\documentclass[pdflatex,sn-mathphys-num]{sn-jnl}

\usepackage{graphicx}
\usepackage{multirow}
\usepackage{amsmath,amssymb,amsfonts}
\usepackage{amsthm}
\usepackage{mathrsfs}
\usepackage[title]{appendix}
\usepackage{xcolor}
\usepackage{textcomp}
\usepackage{manyfoot}
\usepackage{booktabs}
\usepackage{algorithm}
\usepackage{algorithmicx}
\usepackage{algpseudocode}
\usepackage{listings}
\usepackage{diagbox}
\usepackage{bm}
\usepackage{makecell}

\usepackage{array}
\usepackage{textcomp}
\usepackage{epstopdf}
\usepackage{supertabular}
\usepackage{mathtools}
\usepackage{enumitem}

\theoremstyle{thmstyleone}
\newtheorem{defn}{Definition}
\newtheorem{lemma}{Lemma}
\newtheorem{example}{Example}
\newtheorem{theorem}{Theorem}
\newtheorem{remark}{Remark}
\newtheorem{corollary}{Corollary}
\newtheorem{note}{Note}

\newcommand{\floor}[1]{\left\lfloor #1 \right\rfloor}

\newlist{mycases}{enumerate}{1}
\setlist[mycases]{label=\textit{Case~\Roman*:},align=left,itemindent=0pt,leftmargin=0pt,labelwidth=-4pt}

\newlist{mysubcases}{enumerate}{1}
\setlist[mysubcases]{label=\textit{Sub-case~\arabic*:},align=left,itemindent=0pt,leftmargin=0.6cm,labelwidth=-4pt}

\begin{document}
	
	\title[Construction of 2-D Z-Complementary Array Code Sets with Flexible Lengths for Different System Requirements]{Construction of 2-D Z-Complementary Array Code Sets with Flexible Lengths for Different System Requirements}

	\author*[1]{\fnm{Abhishek} \sur{Roy}}\email{1821ma06@iitp.ac.in}
	
	\affil*[1]{\orgdiv{Department of Mathematics}, \orgname{Indian Institute of Technology Patna}, \orgaddress{\street{Bihta}, \city{Patna}, \postcode{801106}, \state{Bihar}, \country{India}}}

	
	\abstract{In this paper, we propose a new and optimal construction of two-dimensional (2-D) Z-complementary array code set (ZCACS) using multivariable extended Boolean functions (EBFs). The proposed 2-D arrays have many applications in modern wireless communications, such as multi-carrier code division multiple access (MC-CDMA), massive multiple input multiple output (mMIMO), etc. The main theoretical problem for sequences and 2-D arrays for application in MC-CDMA lies in the efficient construction of such sequences and arrays, which have low peak-to-mean envelope power ratio (PMEPR) and flexible parameter values. The PMEPR measures the power efficiency of the concerned system and hence has been an important research topic for past several years.
		The proposed construction produces a better PMEPR upper bound than the existing constructions. We also propose a tighter upper bound for the set size which translates more number of supported users in the communication system. We show that for some special cases, the proposed code set is optimal with respect to that bound. Finally, We derive 2-D Golay complementary array set (GCAS) and Golay complementary set (GCS) from the proposed construction, which has significant application in uniform rectangular array (URA)-based massive multiple-input multiple-output (mMIMO) system to achieve omnidirectional transmission. The simulation result shows the performance benefits of the derived arrays. In essence, we show that the flexibility of the parameters of the proposed 2-D ZCACS makes it a good candidate for practical use cases, both in theory and simulation.}

	
	
	\maketitle
	
	\section{Introduction}\label{sec1}
	The area of sequence design with good correlation properties for applications in wireless communication started with the discovery of Golay complementary pairs (GCP), introduced in the seminal paper by Marcel J. E. Golay \cite{GolayInfra}. Leveraging the ideal auto-correlation sum spectrum of these sequences, researchers have found several applications of them in communication systems, for example, channel estimation \cite{ChannelEstm}, radar \cite{DopplerRadar}, reduction of peak-to-mean envelope power ratio (PMEPR) in orthogonal frequency division multiplexing (OFDM) \cite{Popo,DavisJedwab}, etc. However, there are practical limitations on such sequences. For example, a computer search shows that binary GCPs exist only for lengths of the form $2^\alpha 10^\beta 26^\gamma$. To generalize GCP, Tseng \textit{et al.} introduced the Golay complementary set (GCS) in 1972 \cite{TsengCS}, which has increased flexibility in sequence lengths. In the OFDM system, the length of the sequence determines the number of orthogonal subcarriers, and hence is an important parameter. Nevertheless, this flexibility of lengths was achieved with the trade-off in an increased PMEPR, which is upper-bounded by the number of constituent sequences. 
	
	A different direction of generalization was the introduction of zero-correlation zone (ZCZ) by Fan \textit{et al.} \cite{FanZCP}, where the auto-correlation sum is zero in a specific zone around the zero time-shift. However, in a multiple-access system such as multi-carrier code division multiple access (MC-CDMA), an entire sequence set (also sometimes called a code) is assigned to a single user, and hence to support large number of users a large number of such codes are required, which are orthogonal with respect to the correlation function. Research in this direction aims to achieve the optimality of the set size. For this purpose, mutually orthogonal Golay complementary set (MOGCS) \cite{hatoriCCC}, Z-complementary code set (ZCCS) \cite{FanZCP}, low-correlation zone (LCZ) sequence set \cite{LiuLCZ}, and quasi-complementary code set (QCCS)\cite{LiuQCSS} were introduced.
	
	Recently, two-dimensional (2-D) array sets and code sets with their 2-D correlation properties have attracted significant attention due to their practical application benefits, such as ultra wideband (UWB) \cite{Zhang2DMOGCS,XengZCACS}, 2D-MC-CDMA, massive multiple input multiple output (MIMO), etc. The traditional one-dimensional (1-D) correlation functions are special cases of 2-D correlation functions by definition. Hence, it is possible to use 2-D arrays in place of 1-D sequences in multi-carrier systems. The 2-D correlation property of 2-D array sets enables us to control PMEPR in one dimension and achieve interference-free transmission in the other dimension. In such cases, the primary issue is power control, which depends on the correlation property. 
	Turcs{\'a}ny \textit{et al.} in \cite{Turcsny2004New2T,turcsanyQCAS} showed that relaxing the orthogonality property of a 2-D MOGCS, 2-D code sets with larger set size can be used in 2D-MC-CDMA. In such system, the benefit of using 2-D arrays instead of 1-D sequences is the increased freedom to position the array elements in the time and frequency dimensions simultaneously.
	
	Recently, applications of 2-D array as precoders for uniform rectangular array (URA)-based mMIMO \cite{Omni1} has gained significant attention of researchers. The mMIMO system has been extensively used in the 5th generation (5G) of wireless communication. It has the capability to increase the data rate by a large factor in comparison to the single input single output (SISO) system, making it an inseparable feature of future communication standards as well \cite{6Gsurvey}. The literature has many works on 2-D code sets for this application, such as 2-D Golay complementary array sets (GCAS) and 2-D quasi-complementary array sets (QCAS). The corresponding articles can be found here \cite{OmniChen,YuboGCAS,YuboGCAS2,ZilongLiuOmni2024,YuboMIMO2024,YuboQCAS}. 2-D QCAS has an increased flexibility of parameter values in this regard, such as array size and flock size. However, applying 2-D QCAS as precoders has limitations. Because, the power radiation pattern is not omnidirectional in this case, which can be found here \cite{YuboQCAS}. All this suggests that the existing 2-D codes does not completely solve the question of parameter flexibility in omnidirectional transmission. To achieve spatial diversity in such mMIMO system, space-time block code (STBC) is adopted \cite{STBC, OmniChen}, which is used to encode data before transmission. Orthogonal STBCs with lower sizes are suitable for reducing complexity. However, it is an extremely difficult task to achieve the desired flexibility of parameter values for both the URA and STBC  in a single construction of codes, due to the underlying co-dependent relationship between the parameters values. The best solution appears to be finding codes according to the requirements.
	
	In this paper, as the main contribution, we propose 2-D Z-complementary array code sets (ZCACS) of array size $\left( 2^m p \right)  \times \left(  2p_1^{m_1} p_2^{m_2} \dots p_k^{m_k} \right)$, where $m_\alpha \geq 2, \forall \alpha$, $p$, $p_\alpha$'s are prime numbers and $m \geq 1$. To do this, we first propose a construction of inter-group complementary (IGC) code set, which provides the necessary EBFs for the main construction. Then we analyze the PMEPR bounds and show that they are tighter than the existing construction for similar array sizes. We also investigate the set size upper bound and prove a new upper bound for the set size of 2-D ZCACS. It can be seen that the proposed code set is optimal with respect to the bound for $p=2$. Considering the other aspect of application, i.e., mMIMO system, we derive suitable precoding matrices from the 2-D ZCACs to achieve omnidirectional transmission. Simulations are done to show the performance benefit of the precoders and parameter flexibility in URA and STBC.
	
	The rest of the paper is organized as follows: in Section \ref{sec2}, the preliminary definitions and necessary lemmas are discussed. In Section \ref{sec3}, the main construction of 2-D ZCACS is proposed.  In Section \ref{sec4}, the bound for PMEPR is analyzed. In Section \ref{sec5}, the set size bound of 2-D ZCACS is investigated. In Section \ref{sec6}, the application benefit of 2-D code sets in mMIMO is discussed in detail. In Section \ref{sec7}, we compare the proposed construction with previous works with respect to the parameter values. Finally, in Section \ref{sec8}, conclusions are drawn.
	
	\section{Preliminaries}\label{sec2}
	In this section, we discuss some definitions of the sequence/2-D array sets and code sets related to the work. Also, we give a structure of the EBFs used in the construction.
	\subsection{Sequences and 2-D Arrays}
	The sequences and 2-D arrays considered here are assumed to be unimodular, polyphase and complex-valued, which are used in wireless systems with $q$-ary phase shift keying ($q$-PSK) modulation. A 2-D array $\mathbf{X}$ of size $L_1 \times L_2$ is denoted by $X_{i,j}$, where $0 \leq i < L_1$, $0 \leq j < L_2$. For $L_2=1$, it becomes a sequence of length $L_1$, usually denoted by a lowercase bold letter.
	\begin{defn}[2-D ACCF ]
		The 2-D aperiodic cross-correlation function (ACCF) of two arrays $\mathbf{X}$ and $\mathbf{Y}$ of size $(L_1 , L_2)$ at time shift $(\tau_1, \tau_2)$ is defined by
		\begin{equation}
			C (\mathbf{X}, \mathbf{Y})(\tau_1, \tau_2)=
			\sum\limits_{i=0}^{L_1-1} \sum\limits_{j=0}^{L_2-1} X_{i,j} Y^*_{i+\tau_1, j+\tau_2},
		\end{equation}
		where $X_{i,j}, Y_{i,j} =0$ for $i \notin \{0,1,\dots, L_1-1\}$ or $j \notin \{0,1,\dots, L_2-1\}$.
		For $\mathbf{X} = \mathbf{Y}$, it is called 2-D aperiodic auto-correlation function (AACF), denoted by $A(\mathbf{X})(\tau_1, \tau_2)$.
	\end{defn}
	
	\begin{defn}[2-D Array Code]
		An ordered set $\mathcal{X}=\{\mathbf{X}_0, \mathbf{X}_1, \dots, \mathbf{X}_{M-1}\}$ is called a 2-D array code of flock size $M$, where each $\mathbf{X}_\eta$ has size $L_1 \times L_2$.
	\end{defn}
	
	\begin{defn}[2-D GCAS]
		A set of 2-D arrays $\mathcal{X}=\{\mathbf{X}_0, \mathbf{X}_1, \dots, \mathbf{X}_{M-1}\}$ with size $L_1 \times L_2$ is called a 2-D GCAS if they satisfy
		\begin{equation}
			\sum_{\eta=0}^{M-1} A (\mathbf{X}_\eta) (\tau_1, \tau_2)=
			\begin{cases}
				M L_1 L_2, &  (\tau_1, \tau_2)= (0,0);\\
				0, &  \text{otherwise}.
			\end{cases}
		\end{equation}
		For $M=2$, we call it a 2-D Golay complementary array pair (GCAP).
	\end{defn}
	
	\begin{defn}[2-D ZCAC ( \cite{ZengZCACS2})]
		A set of 2-D arrays $\mathcal{X}=\{\mathbf{X}_0, \mathbf{X}_1, \dots, \mathbf{X}_{M-1}\}$ with size $L_1 \times L_2$ is called a 2-D ZCAC if they satisfy
		\begin{equation}
			\sum_{\eta=0}^{M-1} A (\mathbf{X}_\eta) (\tau_1, \tau_2)=
			\begin{cases}
				M L_1 L_2, &  (\tau_1, \tau_2)= (0,0);\\
				0, &  0 \leq \left| \tau_1 \right| < Z_1, 0 \leq \left|\tau_2 \right| < Z_2,\\
				& (\tau_1, \tau_2) \neq (0,0).
			\end{cases}
		\end{equation}
		where $Z_1 \leq L_1$, $Z_2 \leq L_2$. Here, $Z_1 \times Z_2$ is called the rectangular ZCZ size, $M$ is called the flock size.
	\end{defn}
	
	\begin{note}
		When $M = 2$, it is called a 2-D Z-complementary array pair (ZCAP).
	\end{note}
	\begin{defn}[2-D ZCACS (\cite{ZengZCACS2})]
		A collection of 2-D ZCACs $\mathcal{Z}= \{\mathcal{X}_0, \mathcal{X}_1, \dots, \mathcal{X}_{K-1}\}$ where $\mathcal{X}_\nu= \{\mathbf{X}_0^\nu, \mathbf{X}_1^\nu, \dots, \mathbf{X}_{M-1}^\nu\}$ and
		each $\mathbf{X}_i^{\nu}$ has equal size of $L_1 \times L_2$, is called a 2-D ZCACS if they satisfy
		\begin{equation}
			\begin{split}
				\sum_{\eta=0}^{M-1} C (\mathbf{X}_\eta^{\nu_1}, \mathbf{X}_\eta^{\nu_2}) (\tau_1, \tau_2)
				=
				\begin{cases}
					M L_1 L_2, & \nu_1= \nu_2, (\tau_1, \tau_2)= (0,0) ;\\
					0, &  \nu_1= \nu_2, (\tau_1, \tau_2) \neq (0,0),\\
					& 0 \leq \left| \tau_1 \right| < Z_1,  0 \leq  \left| \tau_2 \right| < Z_2;\\
					0, & \nu_1 \neq \nu_2,  \left| \tau_1\right| < Z_1, \left|\tau_2 \right| < Z_2;\\
				\end{cases}
			\end{split}
		\end{equation}
		where $Z_1 \leq L_1$, $Z_2 \leq L_2$. It is denoted as 2-D $(K, Z_1 \times Z_2)$-ZCACS${}_{M}^{L_1 \times L_2}$ and $K$ is called the set size.
	\end{defn}
	
	The known theoretical upper bound on the set size of a 2-D ZCACS is provided below.
	\begin{lemma} [\cite{ZengZCACS2}]\label{lemma_ZCACS_bound}
		For a 2-D $(K, Z_1 \times Z_2)$-ZCACS${}_{M}^{L_1 \times L_2}$,  the following equation holds:
		\begin{equation}
			K  \leq \frac{ M (L_1 + Z_1 -1) (L_2 + Z_2 -1) } {Z_1 Z_2}.
		\end{equation}
		A 2-D ZCACS is called optimal if $K$ achieves its upper bound.
	\end{lemma}
	
	\begin{remark}\label{remark_2D_1D}
		1-D correlation functions and sequence sets appear as a special case of these 2-D correlation functions and 2-D array sets. The corresponding definitions follow analogously only with the change of notations: $L_1=L, L_2=1$ and $Z_1=Z, Z_2=1$. The 1-D counterparts of 2-D GCAS, GCAP, ZCAC, ZCAP, ZCACS are called Golay complementary set (GCS), GCP, Z-complementary set (ZCS), Z-complementary pair (ZCP),  and Z-complementary code set (ZCCS), respectively.
	\end{remark}

	As the paper is concerned with the construction of 2-D ZCACS, we discuss briefly about its 1-D counterpart. We shall see its usefulness later.  The ZCCS with sequence length $L$, ZCZ width $Z$, flock size $M$ and set size $K$ is usually denoted by $(K, Z)$-ZCCS${}_{M}^{L}$. For a 1-D ZCCS we have the following mathematically proved upper bound on the set size.
	\begin{lemma} [\cite{FengZCCSbound}]\label{lemma_ZCCS_bound}
		A $(K, Z)$-ZCCS${}_{M}^{L}$ satisfy
		\begin{equation}
			K  \leq M \left(\frac{L+Z-1}{Z} \right).
		\end{equation}
	\end{lemma}
	
	\begin{note}
		Recently, \cite{pai2024ZCCSbound} proposed a tighter bound, which was a long-standing conjecture. According to it, a $(K, Z)$-ZCCS${}_{M}^{L}$ satisfies  $K \leq ML/Z$. The ZCCS is called optimal when $K = \floor{ML/Z}$, where $\floor{\cdot}$ is the floor function.
	\end{note}

	Now, we define two special sequence sets which have been used in the construction of the proposed 2-D ZCACS.
	
	\begin{defn}[Complementary Mate]
		Two GCPs $(\mathbf{a}_1, \mathbf{b}_1)$ and $(\mathbf{a}_2, \mathbf{b}_2)$ are called complementary mate if they satisfy
		\begin{equation}
			C (\mathbf{a}_1, \mathbf{a}_2) (\tau)+ C (\mathbf{b}_1, \mathbf{b}_2)(\tau)= 0, \forall \tau.
		\end{equation}
	\end{defn}
	
	\begin{defn}[IGC code set (\cite{LiIGC})]
		Consider an optimal $(K,Z)$-ZCCS${}^L_M$ denoted with $\mathcal{S}$ where each ZCS or code is denoted as $\mathcal{C}_{\nu}=\{\mathbf{a}_0^{\nu}, \mathbf{a}_1^{\nu}, \dots, \mathbf{a}_{M-1}^{\nu}\}$.
		Let the codes be divided into $M$ code groups $\mathcal{S}^g$'s. Then $\mathcal{S}$ is called an IGC code set if it satisfies
		\begin{equation}
			\begin{split}
				\sum _{\eta =0}^{M-1} C(\mathbf {a}_\eta^{\nu_1},\mathbf {a}_\eta^{\nu_2})(\tau)=
				\begin{cases} 
					ML, & \nu_1=\nu_2, \tau =0;\\
					0, & \nu_1=\nu_2, 0< \left| \tau \right|<Z;\\ 
					0, & \nu_1\neq \nu_2, \mathcal{C}_{\nu_1},\mathcal{C}_{\nu_2}\in \mathcal{S}^{g}, \left| \tau \right|<Z;\\ 
					0, & \text{otherwise}. 
				\end{cases} \\
			\end{split}	
		\end{equation}
	\end{defn}

	
	\subsection{Extended Boolean Function}	
	In this subsection, we introduce the mathematical tool that has been used for the proposed construction.
	Let $\mathbb{Z}_{n}=\{[0], [1], \dots, [n-1]\}$ be the ring of integers modulo $n$, where $[\cdot]$ denote the residue class. Then any well-defined function $g: \mathbb{Z}_L \rightarrow \mathbb{Z}_q$ produces a sequence of length $L$, namely, $\left(g([0]), g([1]), \dots, g([L-1])\right)$ and it is the fundamental idea behind the sequence generation method using EBFs.
	
	Note that, any length $L$ can be written as $L= p_1^{m_1} p_2^{m_2} \dots p_k^{m_k}$, where $p_\alpha$'s are distinct primes. Then the Chinese Remainder Theorem (CRT) says that $\exists$ a ring isomorphism $\sigma$ such that
	\begin{equation}
		\sigma: \mathbb{Z}_L \cong \mathbb{Z}_{p_1^{m_1}} \times \mathbb{Z}_{p_2^{m_2}} \times \dots \times \mathbb{Z}_{p_k^{m_k}},
	\end{equation}
	given by $x \mapsto \left(x \mod p_1^{m_1}, x \mod p_2^{m_2}, \dots, x \mod p_k^{m_k} \right)$.
	
	Now, for the purpose of maintaining the well-defined property of EBFs, we assume $\mathbb{Z}_n$ as the set $\{0,1, \dots, n-1\}$ of numbers only, dropping the residue-class property. The corresponding functions rules are set accordingly to map the elements of $\mathbb{Z}_L$ to $\mathbb{Z}_q$, which we show shortly.
	
	We can define a bijection $\varphi: \mathbb{Z}_{p_\alpha^{m_\alpha}} \rightarrow \mathbb{Z}_{p_\alpha}^{m_\alpha}$ given by $\varphi (x)= \vec{x}= (x_1, x_2, \dots, x_{m_\alpha})$, where $\vec{x}$ is the $p_\alpha$-ary representation of the integer $x= \sum_{i=1}^{m_\alpha} x_i p_\alpha^{i-1}$. Hence, we get a bijective map
	\begin{equation}\label{eqn_CRT_comb}
		\left( \varphi \circ \sigma \right): \mathbb{Z}_L \rightarrow \mathbb{Z}_{p_1}^{m_1} \times \mathbb{Z}_{p_2}^{m_2} \times \dots \times \mathbb{Z}_{p_k}^{m_k}.
	\end{equation}
	
	Now, for the co-domain of EBFs, we first take the set $\mathbb{Z}_q$ for any $q$. Then an EBF $f: \mathbb{Z}_L \rightarrow \mathbb{Z}_q$ can be thought of as a function
	\begin{equation}
		f': \mathbb{Z}_{p_1}^{m_1} \times \mathbb{Z}_{p_2}^{m_2} \times \dots \times \mathbb{Z}_{p_k}^{m_k} \rightarrow \mathbb{Z}_q,
	\end{equation}
	where $f(x)= f' \left((\varphi \circ \sigma)(x) \right)$ and $f'$ can be written as a function of $m'=\sum_{\alpha=1}^{k}m_\alpha$ variables $\{v_1, v_2, \dots, v_{m'}\}$. To make the function well-defined, we adopt the function rule as
	\begin{equation}
		\left(x_1, x_2, \dots, x_{m'}\right) \mapsto f'_{\mathbb{Z}} (x_1, x_2, \dots, x_{m'}) \mod q,
	\end{equation}
	where $f'_{\mathbb{Z}} (x_1, x_2, \dots, x_{m'})$ denotes the value of the polynomial expression of $f'(x_1, x_2, \dots, x_{m'})$ in $\mathbb{Z}$.
	
	\begin{remark}
		Note that, if we apply a permutation function $\phi$ on $\sigma$, then the resultant function
		\begin{equation}
			\left( \varphi \circ (\phi \circ \sigma) \right) : \mathbb{Z}_L \rightarrow \mathbb{Z}_{p_1}^{m_1} \times \mathbb{Z}_{p_2}^{m_2} \times \dots \times \mathbb{Z}_{p_k}^{m_k}
		\end{equation}
		still offers a valid representations of the numbers $0 \leq i < p_1^{m_1} p_2^{m_2} \dots p_k^{m_k}$, with some permutation on the original representation $\left( \varphi \circ \sigma \right) (i)$. For sequence generation in this paper, we shall need such permuted representation of $i$.
		This definition of EBFs allows us to permute the components in the direct product, i.e., $\mathbb{Z}_{p_\alpha}$'s, so that we have a more general form
		\begin{equation}\label{eq_ebf2}
			f' : \mathbb{Z}_{p'_1}^{m'_1} \times \mathbb{Z}_{p'_2}^{m'_2} \times \dots \times \mathbb{Z}_{p'_k}^{m'_k} \rightarrow \mathbb{Z}_q,
		\end{equation}
		where $p'_\alpha$'s may not be distinct. We shall use this general form.
	\end{remark}

	For the construction of 2-D ZCACS, we need the co-domain $\mathbb{Z}_q$ to be extended, which will allow some multiplicative invertible elements to appear in the range. In this process, we show that different kinds of such extensions are possible. 
	\begin{enumerate}
		\item When considering $\mathbb{Z}_q$ as a ring, we may take the set $D= \{x: x \in \mathbb{Z}_q, x^n \neq 0, \forall n \in \mathbb{N}\}$ such that $D$ is closed with respect to multiplication, and no nilpotent element exists in $D$. Then, we may construct the ring of fractions $D^{-1} \mathbb{Z}_q = \{x d^{-1}: x \in \mathbb{Z}_q, d \in D\}$ by using localization, so that $\mathbb{Z}_q \subseteq D^{-1} \mathbb{Z}_q$ as a subring. This extension inherently provides well-defined binary operations in $D^{-1} \mathbb{Z}_q$. Note that such a $D$ always exists in $\mathbb{Z}_q$.
		
		\item Or, we may consider the set $\mathbb{Z}_q=\{0,1, \dots, q-1\}$ as a set of integers only. Now, we can define a quotient set $\mathbb{Q}_q=\left\{\frac{a}{b} \mid a, b \in \mathbb{Z}_q, b \neq 0\right\}$, where the addition and multiplication rule are set in such a way that they are well-defined and it avoids the degenerate case, i.e., $\mathbb{Q}_q= \{0\}$. 
		This can be done by defining 
		\begin{equation}
			(x_1, x_2, \dots, x_{m'}) \mapsto f'_{\mathbb{Q}} (x_1, x_2, \dots, x_{m'})
		\end{equation}
		where $f'_{\mathbb{Q}} (x_1, x_2, \dots, x_{m'}):= \frac{a_{f'} \mod q}{b_{f'} \mod q}$, $a_{f'}, b_{f'} \in \mathbb{Z}$. If it occurs that 
		$b_{f'} \mod q = 0 (\mod q)$, then we may define $f'_{\mathbb{Q}} (x_1, x_2, \dots, x_{m'}):= \kappa$, for some $\kappa \in \mathbb{Q}_q$.
	\end{enumerate}
	For the sake of simplicity, we denote both types of extensions as $\mathbb{Q}_q$ and take either extension. It does not matter in the actual construction process, as we shall see later.
	
	From (\ref{eq_ebf2}), we derive the associated complex-valued sequence corresponding to $f'$.
	Let $v_{p'_\alpha,1}, v_{p'_\alpha,2}, \dots, v_{p'_\alpha,m_\alpha}$ denote the $m_\alpha$ variables which take values from $\mathbb{Z}_{p'_\alpha}$, $\forall \alpha$. The radix-representation of $i$ is defined as $\left( \vec{i}_1, \vec{i}_2, \dots, \vec{i}_k \right)$.
	We can associate an EBF $f$ with a $\mathbb{Q}_q$-valued sequence of length $p_1^{m_1} p_2^{m_2} \dots p_k^{m_k}$ using
	\begin{equation}
		\textbf{f}= \bigg( f \left( \vec{i}_1, \vec{i}_2, \dots, \vec{i}_k \right) :  0 \leq i < {p'}_1^{m_1} {p'}_2^{m_2} \cdots {p'}_k^{m_k} \bigg),
	\end{equation}
	where $i$ is varied in the following manner: 
	\textit{First $\vec{i}_1$ is varied from $0$ to ${p'}_1^{m_1}-1$, keeping others fixed at zero-vector. Then both $\vec{i}_2$ and $\vec{i}_1$ are varied, keeping others fixed. In summary, at $l$-th step $\{\vec{i}_{l}, \dots, \vec{i}_1\}$ are varied, keeping the rest fixed, for $l=1,2, \dots, k$. This gives an enumeration of the integers $0 \leq i < {p'}_1^{m_1} {p'}_2^{m_2} \dots {p'}_k^{m_k}$ similar to a Boolean-representation table.}
	
	One can also associate a complex-valued sequence corresponding to an EBF $f$, which is given by
	\begin{equation}
		\psi(f)= \bigg(\omega_q^{ f\left( \vec{i}_1, \vec{i}_2, \dots, \vec{i}_k \right) } :  0 \leq i < {p'}_1^{m_1} {p'}_2^{m_2} \cdots {p'}_k^{m_k} \bigg),
	\end{equation}
	where $\omega_q= \exp\left(\frac{2\pi \sqrt{-1}}{q}\right)$.
	
	\begin{note}
		For $p_\alpha=2, \forall \alpha$, this definition reduces to the definition of generalized Boolean function (GBF) $f: \mathbb{Z}_2^m \rightarrow \mathbb{Z}_q$.
	\end{note}
	
	\subsection{2-D Extended Boolean Function}
	We further introduce some notations to produce 2-D arrays by using multivariable EBFs. Let $F_{\vec{d}}: \mathbb{Z}_{p_1}^{m_1} \times \mathbb{Z}_{p_2}^{m_2} \times \dots \times \mathbb{Z}_{p_k}^{m_k} \rightarrow \mathbb{Q}_q$ be a set of EBFs for $\vec{d}= (d_{p'_1}, d_{p'_2}, \dots, d_{p'_w}) \in \prod_{\alpha=1}^{w} \mathbb{Z}_{p'_\alpha}$, where $p'_\alpha$ are some distinct primes and $\vec{d}$ is the radix-representation of number $d$ for $0 \leq d < p'_1 p'_2 \cdots p'_k$. Let $K = \{F_{ \vec{d}}: \vec{d} \in \prod_{\alpha=1}^{w} \mathbb{Z}_{p'_\alpha} \}$ be an ordered set, which can be expressed as a $\mathbb{Q}_q$-valued 2-D array of size $L_1 \times L_2$ by writing the corresponding sequences, where $L_1= p'_1p'_2 \dots p'_w$ and $L_2= p_1^{m_1} p_2^{m_2} \dots p_k^{m_k}$. The $d$-th row of $K$ is determined by the EBF $F_{ \vec{d}}$. 
	
	Similar to the 1-D case, we can also associate a complex-valued matrix $\Psi(K)$ with $K$, where $\left( \Psi(K) \right)_{i,j}= \omega_q^{K_{i,j}}$ for all $i,j$. Below, we show an example of this for ease of understanding.

	\begin{example}
		Let $F_{ \vec{d}}: \mathbb{Z}_{2} \times \mathbb{Z}_{3}  \rightarrow \mathbb{Q}_4$ be defined by  $F_{ \vec{d}} (v_{2,1}, v_{3,1})= 2 v_{2,1} d_2 + 3 v_{3,1} d_3$, where $\vec{d}= (d_2, d_3)$. Then the $6 \times 6$ 2-D array $K$ related to $F_{\vec{d}}$ is generated in Table \ref{table_array_generate}.
	\end{example}
	
	\begin{table}[h]
		\caption{2-D array corresponding to the functions $F_{\vec{d}}$ }
		\label{table_array_generate}
		\centering
		\begin{tabular}{|c|cccccc|}
			\hline
			\diagbox{$\vec{d}= (d_2, d_3)$}{ $(v_{2,1}, v_{3,1}$)} & (0,0)& (1,0)& (0,1)& (1,1)& (0,2)& (1,2)\\ 
			\hline
			(0,0)& 0& 0& 0& 0& 0& 0\\ 
			(1,0)& 0& 2& 0& 2& 0& 2\\ 
			(0,1)& 0& 0& 3& 3& 2& 2\\ 
			(1,1)& 0& 2& 3& 1& 2& 0\\ 
			(0,2)& 0& 0& 2& 2& 0& 0\\ 
			(1,2)& 0& 2& 2& 0& 0& 2\\ 
			\hline
		\end{tabular}
	\end{table}
	
	
	\subsection{Peak-to-Mean Envelope Power Ratio}
	In this subsection, we discuss the problem of PMEPR reduction in multi-carrier systems. In a MC-CDMA system, column sequence of 1-D code is considered for power control \cite{LiuMCCDMA}. For a 2-D code, a 2-D array of that code can also be used, using any one of the dimensions for power control. Given a $\mathbb{Q}_q$-valued sequence $\mathbf{a}= (a_0, a_1, \dots, a_{L-1} )$, the transmitted signal is modeled as the real part of the complex envelope
	\begin{equation}
		S_{\mathbf{a}} (t)= \sum_{i=0}^{L-1} \omega_q^{a_i + q f_i t},
	\end{equation}
	where  $f_i= f + i \Delta f$, $f$ is a constant, $\Delta f$ is an integer multiple of symbol rate and $0 \leq \Delta f t \leq 1$ \cite{DavisJedwab}.  The term $\left|  S_{\mathbf{a}} (t)  \right|^2/L$ is called the instantaneous-to-average power ratio (IAPR).
	The PMEPR of the sequence $\mathbf{a}$ is given by
	\begin{equation}
		\text{PMEPR}(\mathbf{a})= \sup \limits_{0 \leq \Delta f t \leq 1} \frac{\mid S_{\mathbf{a}} (t) \mid ^2 }{L}.
	\end{equation}
	In an ideal case, PMEPR should be as low as possible.

	
	\subsection{mMIMO-based Omnidirectional Beamforming}
	In this subsection, discuss the application of 2-D codes in mMIMO with URA for omnidirectional signal transmission.  As described in \cite{Omni1, Omni2}, we consider a downlink transmission from a base station (BS) equipped with URA to a single-antenna user equipment (UE).  We consider a $P \times Q$ URA of $PQ$ antennas. Let, $A( \phi, \theta)$ be the steering matrix corresponding to the URA at the direction angle $(\phi, \theta)$, where $\phi \in [0, \pi/2]$ and $\theta \in [0, 2 \pi]$. Then $A(\phi, \theta)$ is given by
	\begin{equation}\label{steering}
		\begin{split}
			[A(\phi, \theta)]_{i,j}= e^{-\iota \frac{2 \pi}{\lambda} i d_y \sin \phi \sin \theta - \iota \frac{2 \pi}{\lambda} j d_x \sin \phi \cos \theta},
		\end{split}
	\end{equation}
	where $\iota= \sqrt{-1}$, $0 \leq i \leq P-1$, $0 \leq j \leq Q-1$,  $\lambda$ is the wavelength of the carrier, $d_x$ and $d_y$ are the inter-element distance of URA along the vertical and horizontal directions, respectively. At first, the data bits are modulated into a $K \times N$-STBC given by $\mathbf{S}$, where the $(n,t)$-th entry is $s_n(t)$, for $n=0,1,\dots, K-1$ and $t=0,1,\dots, N-1$ \cite{STBC}. It is then beamformed with the precoding matrices $\mathbf{W}_n$ of size $P \times Q$ and sent over the wireless channel. The received signal in a line-of-sight (LOS) without multi-paths is given by
	\begin{equation}\label{received_signal}
		y(t)= \sum_{n=0}^{K-1} \left[vec(A(\phi, \theta))^T \cdot vec(\mathbf{W}_n)\right] s_n(t) + w(t),
	\end{equation}
	for $0 \leq t \leq N-1$,
	where $vec(\cdot)$ denotes the vector produced by vertically stacking the columns of the corresponding matrix,  $(\cdot)^T$ denotes the transpose of the matrix $(\cdot)$, and $w(t)$ is the additive white Gaussian noise (AWGN). 
	
	The next two lemmas show the usefulness of 1-D and 2-D codes used in such system for omnidirectional power property.
	\begin{lemma}[\cite{Omni2}]\label{lemma_GCAS_omni}
		Let the precoding matrices $\{\mathbf{W}_0, \mathbf{W}_1, \dots, \mathbf{W}_{K-1}\}$ of size $P \times Q$ constitute a 2-D GCAS. Then the received power $\sum_{n=0}^{K-1} \left|  \left[vec(A(\phi, \theta))^T \cdot vec(\mathbf{W}_n)\right] \right|^2$ is independent of the direction $(\phi, \theta)$. 
	\end{lemma}
	\begin{lemma}[\cite{Omni1}]\label{lemma_CS_omni}
		Let $\{\mathbf{c}_0, \mathbf{c}_1, \dots, \mathbf{c}_{P-1}\}$ be a GCS of length $Q$, with $P=K$. Then the received power $\sum_{n=0}^{K-1} \left|  \left[vec(A(\phi, \theta))^T \cdot vec(\mathbf{W}_n)\right] \right|^2$ is independent of the direction $(\phi, \theta)$ if $vec(\mathbf{W}_n)$ is chosen to be
		\begin{equation}
			vec(\mathbf{W}_n)= \left[  \underbrace{\mathbf{0} \cdots \mathbf{0}}_{n-\text{times}}~~~  \mathbf{c}_n  \underbrace{\mathbf{0} \cdots \mathbf{0}}_{(P-n-1)-\text{times}} \right]^{T},
		\end{equation}
		where $\mathbf{0}$ denotes the zero matrix of size $1 \times Q$.
	\end{lemma}

	\section{Construction of 2-D ZCACS}\label{sec3}
	In this section, we propose the construction of 2-D ZCACS along with its constituent 2-D codes which is 2-D ZCAC. However, for this purpose, we first state a lemma that produces IGC code sets of flexible lengths of the form $p_1^{m_1-1} p_2^{m_2-1} \dots p_k^{m_k-1} (m_\alpha \geq1, \forall \alpha)$.

	\subsection{IGC Code Set}
	We take the EBFs
	$f_\alpha : \mathbb{Z}_{p_\alpha}^{m_\alpha-1} \rightarrow \mathbb{Q}_q$, 
	for $\alpha = 1, 2, \dots, k$, where each $f_\alpha$ is defined by
	
	\begin{equation}
		\begin{split}
			f_\alpha
			&= \frac{q}{p_\alpha}  \sum_{\beta=1}^{m_\alpha-2}  v_{p_\alpha, \pi_\alpha(\beta)} v_{p_\alpha, \pi_\alpha(\beta+1)} +  \sum_{\beta=1}^{m_\alpha-1}  c_{\alpha, \beta} v_{p_\alpha, \beta} + c_\alpha,
		\end{split}
	\end{equation}
	where $c_{\alpha, \beta}, c_\alpha \in \mathbb{Q}_q$, $m_\alpha \geq 2$ for all $\alpha$, $\pi_\alpha$ is a permutation of the set $\{1,2, \dots, m_\alpha-1\}$, and $p_\alpha \mid q$ for all $\alpha$.
	Let the function
	\begin{equation}
		f: \mathbb{Z}_{p_1}^{m_1-1} \times \mathbb{Z}_{p_2}^{m_2-1} \times \dots \times \mathbb{Z}_{p_k}^{m_k-1} \rightarrow \mathbb{Q}_q
	\end{equation} 
	be defined by
	\begin{equation}
		f= f_1 + f_2+ \dots + f_k.
	\end{equation}
	Now, for $\bm{\gamma}= (\gamma_1, \gamma_2, \dots, \gamma_k),\mathbf{s}= (s_1, s_2, \dots, s_k), \mathbf{t}=(t_1, t_2, \dots, t_k) \in \mathbb{Z}_{p_1} \times \mathbb{Z}_{p_2} \times \dots \times \mathbb{Z}_{p_k}$, we define the set of functions
	\begin{equation}
		a_{\mathbf{s},\mathbf{t}}^{\bm{\gamma}}: \mathbb{Z}_{p_1}^{m_1-1} \times \dots \times \mathbb{Z}_{p_k}^{m_k-1} \times \mathbb{Z}_{p_1} \times \dots \times \mathbb{Z}_{p_k} \rightarrow \mathbb{Q}_q,
	\end{equation} 
	given by 
	\begin{equation}
		a_{\mathbf{s},\mathbf{t}}^{\bm{\gamma}}=  R_{\mathbf{t}}^{\bm{\gamma}} + T_{\mathbf{s}},
	\end{equation}
	where $ R_{\mathbf{t}}^{\bm{\gamma}} :  \mathbb{Z}_{p_1}^{m_1-1} \times \dots \times \mathbb{Z}_{p_k}^{m_k-1} \rightarrow \mathbb{Q}_q$ is defined by
	
	\begin{equation}
		R_{\mathbf{t}}^{\bm{\gamma}} = f+ \sum_{\alpha=1}^{k} \frac{q}{p_\alpha} v_{p_\alpha, \pi_\alpha(1)} \gamma_\alpha + \sum_{\alpha=1}^{k} \frac{q}{p_\alpha} v_{p_\alpha, \pi_\alpha(m_\alpha-1)} t_\alpha,
	\end{equation}
	and 
	$ T_{\mathbf{s}} :  \mathbb{Z}_{p_1} \times \dots \times \mathbb{Z}_{p_k} \rightarrow \mathbb{Q}_q$ is defined by
	
	\begin{equation}
		T_{\mathbf{s}} = \sum_{\alpha=1}^{k} \frac{q}{p_\alpha} v'_{p_\alpha} s_\alpha.
	\end{equation}
	Here, $v'_{p_\alpha}$'s are the variables corresponding to the last $\mathbb{Z}_{p_\alpha}$'s.
	Next, we define the code $\mathcal{C}_{\mathbf{s}, \mathbf{t}}$ which is given by
	
	\begin{equation}
		\mathcal{C}_{\mathbf{s}, \mathbf{t}}= 
		\begin{bmatrix}
			\vdots\\
			\psi( a_{\mathbf{s},\mathbf{t}}^{\bm{\gamma}} )\\
			\vdots\\
		\end{bmatrix}_{\bm{\gamma} \in \mathbb{Z}_{p_1} \times \mathbb{Z}_{p_2} \times \dots \times \mathbb{Z}_{p_k}},
	\end{equation}
	which is a $M \times L$ matrix with $M= p_1 p_2 \dots p_k$ and $L= p_1^{m_1} p_2^{m_2} \dots p_k^{m_k}$. The next lemma provides the construction of the IGC code set.

	\begin{lemma}\label{Thm_IGC_gen}
		The code set $\mathcal{S} = \{ \mathcal{C}_{\mathbf{s}, \mathbf{t}}: \mathbf{s}, \mathbf{t} \in \prod_{\alpha=1}^{k} \mathbb{Z}_{p_\alpha}\}$  forms a $(K,M, L,Z)$-IGC code set with $K= (p_1 p_2 \dots p_k)^2$, $M= p_1 p_2 \dots p_k$, $L= p_1^{m_1} p_2^{m_2} \dots p_k^{m_k}$ and $Z= p_1^{m_1-1} p_2^{m_2-1} \dots p_k^{m_k-1}$. The $\mathbf{t}$-th IGC code group is expressed as $I^{\mathbf{t}} = \{ \mathcal{C}_{\mathbf{s}, \mathbf{t}}: \mathbf{s} \in \prod_{\alpha=1}^{k} \mathbb{Z}_{p_\alpha}\}$.
	\end{lemma}

	We state two simple lemmas which will be useful for the proof of Lemma \ref{Thm_IGC_gen}. The lemma follows directly from the construction of MOGCS in \cite{sarkar2021multivariable}. The second lemma is proved after its statement.
	
	\begin{lemma}[\cite{sarkar2021multivariable}]\label{lemma_CCC}
		For $\mathbf{t}_1, \mathbf{t}_2 \in \mathbb{Z}_{p_1} \times \mathbb{Z}_{p_2} \times \dots \times \mathbb{Z}_{p_k}$, we have
		
		\begin{equation}\label{Lemma_1_eqn}
			\begin{split}
				\sum_{\bm{\gamma \in \prod_{\alpha=1}^{k} \mathbb{Z}_{p_\alpha}}}^{} C\left( \psi (R_{\mathbf{t}_1}^{\bm{\gamma}} ), \psi (R_{\mathbf{t}_2}^{\bm{\gamma}} ) \right) (\tau)
				= 
				\begin{cases}
					\prod_{\alpha=1}^{k} p_\alpha^{m_\alpha}, & \mathbf{t}_1 = \mathbf{t}_2, \tau=0;\\
					0, & \mathbf{t}_1 = \mathbf{t}_2, 0< \mid \tau \mid < \prod_{\alpha=1}^{k} p_\alpha^{m_\alpha-1};\\
					0, & \mathbf{t}_1 \neq \mathbf{t}_2, \mid \tau \mid < \prod_{\alpha=1}^{k} p_\alpha^{m_\alpha-1}.
				\end{cases}
			\end{split}
		\end{equation}
	\end{lemma}

	\begin{lemma}\label{lemma_hadamard}
		For $\mathbf{s}_1, \mathbf{s}_2 \in \mathbb{Z}_{p_1} \times \mathbb{Z}_{p_2} \times \dots \times \mathbb{Z}_{p_k}$, we have
		\begin{equation}\label{Lemma_2_eqn}
			C \left( \psi( T_{\mathbf{s}_1}), \psi( T_{\mathbf{s}_2}) \right) (0) =
			\begin{cases}
				\prod_{\alpha=1}^{k} p_\alpha, & \mathbf{s}_1 = \mathbf{s}_2;\\
				0, & \mathbf{s}_1 \neq \mathbf{s}_2.
			\end{cases}
		\end{equation}
	\end{lemma}
	\begin{proof}
		We let $\mathbf{s}_i= (s^i_1, s^i_2, \dots, s^i_k)$ for $i=1,2$.
		From the definition of $T_\mathbf{s}$, we have
		\begin{equation}\label{lemma_hadamard_eqn}
			\begin{split}
				C \left( \psi( T_{\mathbf{s}_1}), \psi( T_{\mathbf{s}_2}) \right) (0)
				= &\sum_{v'_{p_1} = 0}^{p_1-1}   \sum_{v'_{p_2} = 0}^{p_2-1} \dots \sum_{v'_{p_k} = 0}^{p_k-1} \omega_q^{\sum_{\alpha=1}^{k} \frac{q}{p_\alpha} v'_{p_\alpha} (s^1_\alpha- s^2_\alpha)}\\
				= &\sum_{v'_{p_1} = 0}^{p_1-1} \omega_{p_1}^{  v'_{p_1} (s^1_1- s^2_1)} \times \dots \times \sum_{v'_{p_k} = 0}^{p_k-1} \omega_{p_k}^{  v'_{p_k} (s^1_k- s^2_k)}.
			\end{split}			
		\end{equation}
		When $\mathbf{s}_1 = \mathbf{s}_2$, then $s^1_\alpha= s^2_\alpha, \forall \alpha$. So, we have
		\begin{equation}\label{lemma_hadamard_eqn1}
			\sum_{v'_{p_\alpha} = 0}^{p_\alpha-1} \omega_{p_\alpha}^{  v'_{p_\alpha} (s^1_\alpha- s^2_\alpha)}= p_\alpha, \forall \alpha.
		\end{equation}
		From (\ref{lemma_hadamard_eqn}) and (\ref{lemma_hadamard_eqn1}), for $\mathbf{s}_1 = \mathbf{s}_2$ we have
		\begin{equation}
			C \left( \psi( T_{\mathbf{s}_1}), \psi( T_{\mathbf{s}_2}) \right) (0) =  \prod_{\alpha=1}^{k} p_{\alpha}.
		\end{equation}
		On the other hand, when $\mathbf{s}_1 \neq \mathbf{s}_2$, then $s^1_\delta \neq s^2_\delta$ for some $\delta \in \{1,2, \dots, k\}$. So, we have
		\begin{equation}\label{lemma_hadamard_eqn2}
			\sum_{v'_{p_\delta} = 0}^{p_\delta-1} \omega_{p_\delta}^{  v'_{p_\delta} (s^1_\delta- s^2_\delta)}=0,
		\end{equation}
		because $\{\omega_{p_\delta}^{  v'_{p_\delta} (s^1_\delta- s^2_\delta)}:  v'_{p_\delta} =0,1, \dots,p_\delta-1\}$ are all the roots of the polynomial $z^{p_\delta}-1$.
		Hence, from (\ref{lemma_hadamard_eqn}) and (\ref{lemma_hadamard_eqn2}), for  $\mathbf{s}_1 \neq \mathbf{s}_2$ we have the result.
	\end{proof}
	
	Now, we can prove Lemma \ref{Thm_IGC_gen}. We consider the following observation before proving the lemma. Note that for two pairs of sequences $(\mathbf{a}_1, \mathbf{a}_2)$, $(\mathbf{b}_1, \mathbf{b}_2)$ and $\tau \geq 0$, we have
	\begin{equation}\label{eqn_Kronecker}
		\begin{split}
			C (  \mathbf{a}_1 \otimes \mathbf{b}_1, \mathbf{a}_2 \otimes \mathbf{b}_2   ) (\tau)= &C ( \mathbf{a}_1, \mathbf{a}_2) \left( \floor{\frac{\tau}{N}}\right) C ( \mathbf{b}_1, \mathbf{b}_2 ) ( \tau  \mod N)\\
			&+  \Delta_N C ( \mathbf{a}_1, \mathbf{a}_2)  \left( \floor{\frac{\tau}{N}}  +  1 \right)  C ( \mathbf{b}_1, \mathbf{b}_2 ) ( \tau  \mod N - N),
		\end{split}
		\raisetag{5mm}
	\end{equation}
	where $\otimes$ denotes the Kronecker product, each $\mathbf{a}_i$ has length $M$ and each $\mathbf{b}_i$ has length $N$, and 
	\begin{equation}
		\Delta_N=
		\begin{cases}
			0, &\tau \mod N=0;\\
			1, &\text{otherwise}.
		\end{cases}
	\end{equation}

	\begin{proof}[Proof of Lemma \ref{Thm_IGC_gen}]
		We can write $\psi(a_{\mathbf{s},\mathbf{t}}^{\bm{\gamma}} )$ as
		\begin{equation}
			\psi(a_{\mathbf{s},\mathbf{t}}^{\bm{\gamma}} )  = \psi( T_{\mathbf{s}}) \otimes \psi (R_{\mathbf{t}}^{\bm{\gamma}} ).
		\end{equation} 
		Then we have
		\begin{equation}\label{eqn_Kronecker_detailed}
			\begin{split}
				& \sum_{\bm{\gamma} \in \prod_{\alpha=1}^{k} \mathbb{Z}_{p_\alpha}}^{} C\left( \psi(a_{\mathbf{s}_1,\mathbf{t}_1}^{\bm{\gamma}} ), \psi(a_{\mathbf{s}_2,\mathbf{t}_2}^{\bm{\gamma}} ) \right) (\tau)\\
				&= \bigg[ C\left( \psi( T_{\mathbf{s}_1}), \psi( T_{\mathbf{s}_2}) \right) \left( \floor{\frac{\tau}{N}}\right) \sum_{\bm{\gamma} \in \prod_{\alpha=1}^{k} \mathbb{Z}_{p_\alpha}}^{}  C \left( \psi (R_{\mathbf{t}_1}^{\bm{\gamma}} ), \psi (R_{\mathbf{t}_2}^{\bm{\gamma}} ) \right) \left(\tau  \mod N \right) \bigg]\\
				& \quad + \bigg[\Delta_N C\left( \psi( T_{\mathbf{s}_1}), \psi( T_{\mathbf{s}_2}) \right) \left( \floor{\frac{\tau}{N}} +1\right)  \times \sum_{\bm{\gamma} \in \prod_{\alpha=1}^{k} \mathbb{Z}_{p_\alpha}}^{} \!\!\!\!\!\!\! C \left( \psi (R_{\mathbf{t}_1}^{\bm{\gamma}} ), \psi (R_{\mathbf{t}_2}^{\bm{\gamma}} ) \right) \left(\tau \mod N -N\right) \bigg],
			\end{split}
		\end{equation}
		where $N= \prod_{\alpha=1}^{k} p_\alpha^{m_\alpha-1}$.
		We need to show that
		\begin{equation}\label{Thm_2_main_eqn_1}
			\begin{split}
				\sum_{\bm{\gamma} \in \prod_{\alpha=1}^{k} \mathbb{Z}_{p_\alpha}}^{} C\left( \psi(a_{\mathbf{s}_1,\mathbf{t}_1}^{\bm{\gamma}} ), \psi(a_{\mathbf{s}_2,\mathbf{t}_2}^{\bm{\gamma}} ) \right) (\tau)
				=
				\begin{cases}
					\prod_{\alpha=1}^{k} p_\alpha^{m_\alpha+1}, & \left( \mathbf{s}_1, \mathbf{t}_1 \right) = \left( \mathbf{s}_2, \mathbf{t}_2 \right),  \tau=0;\\
					0, & \left( \mathbf{s}_1, \mathbf{t}_1 \right) = \left( \mathbf{s}_2, \mathbf{t}_2 \right), \\
					&0 < \tau < \prod_{\alpha=1}^{k} p_\alpha^{m_\alpha-1};\\
					0, & \mathbf{s}_1 \neq \mathbf{s}_2, \mathbf{t}_1 =\mathbf{t}_2,\\
					& \tau < \prod_{\alpha=1}^{k} p_\alpha^{m_\alpha-1}.\\
				\end{cases}
			\end{split}
		\end{equation}
		Here we considered only $\tau \geq 0$, because $C(\mathbf{a}, \mathbf{b})(-\tau ) = C^*(\mathbf{b}, \mathbf{a})(\tau )$.
		The rest of the proof is split into three cases.
		\begin{mycases}
			\item Let $\tau = 0$. Then $\floor{\frac{\tau}{N}}=0$, $\tau \mod N = 0$ and $\Delta_N=0$. Using Lemma \ref{lemma_CCC}, Lemma \ref{lemma_hadamard}, and (\ref{eqn_Kronecker_detailed}), we have
			\begin{equation}\label{Thm_2_case_1_eqn_1}
				\begin{split}
					\sum_{\bm{\gamma} \in \prod_{\alpha=1}^{k} \mathbb{Z}_{p_\alpha}}^{} C\left( \psi(a_{\mathbf{s}_1,\mathbf{t}_1}^{\bm{\gamma}} ), \psi(a_{\mathbf{s}_2,\mathbf{t}_2}^{\bm{\gamma}} ) \right) (\tau)
					=&
					\begin{cases}
						\prod_{\alpha=1}^{k} p_\alpha^{m_\alpha+1}, & \mathbf{s}_1= \mathbf{s}_2, \mathbf{t}_1 =\mathbf{t}_2;\\
						0, & \mathbf{s}_1 \neq \mathbf{s}_2, \mathbf{t}_1 =\mathbf{t}_2;\\
						0, & \mathbf{t}_1 \neq \mathbf{t}_2.
					\end{cases}
				\end{split}
			\end{equation}
			\item Let $0 < \tau < \prod_{\alpha=1}^{k} p_\alpha^{m_\alpha-1}$. In this case $\floor{\frac{\tau}{N}}=0$, $0 < \tau \mod N  < \prod_{\alpha=1}^{k} p_\alpha^{m_\alpha-1}$, $-\prod_{\alpha=1}^{k} p_\alpha^{m_\alpha-1}< \tau \mod N-N <0 $ and $\Delta_N=1$.
			Using \textit{Lemma \ref{lemma_CCC}}, we have
			\begin{equation}\label{Thm_2_case_2_eqn_1}
				\begin{split}
					\sum_{\bm{\gamma} \in \prod_{\alpha=1}^{k} \mathbb{Z}_{p_\alpha}}^{} C \left( \psi (R_{\mathbf{t}_1}^{\bm{\gamma}} ), \psi (R_{\mathbf{t}_2}^{\bm{\gamma}} ) \right) \left(\tau \mod N \right)= &0,\\
					\sum_{\bm{\gamma} \in \prod_{\alpha=1}^{k} \mathbb{Z}_{p_\alpha}}^{} C \left( \psi (R_{\mathbf{t}_1}^{\bm{\gamma}} ), \psi (R_{\mathbf{t}_2}^{\bm{\gamma}} ) \right) \left(\tau \mod N-N \right)= &0,
				\end{split}
			\end{equation}
			for all values of $\mathbf{t}_1$ and $\mathbf{t}_2$. Then, from (\ref{eqn_Kronecker_detailed}) and (\ref{Thm_2_case_2_eqn_1}), for all choices of $\left( \mathbf{s}_1, \mathbf{t}_1\right)$ and $\left( \mathbf{s}_2, \mathbf{t}_2\right)$, we have
			\begin{equation}
				\begin{split}
					\sum_{\bm{\gamma} \in \prod_{\alpha=1}^{k} \mathbb{Z}_{p_\alpha}}^{}  C\left( \psi(a_{\mathbf{s}_1,\mathbf{t}_1}^{\bm{\gamma}} ), \psi(a_{\mathbf{s}_2,\mathbf{t}_2}^{\bm{\gamma}} ) \right) (\tau)=0.
				\end{split}
			\end{equation}
			\item Let $\prod_{\alpha=1}^{k} p_\alpha^{m_\alpha-1} \leq \tau < \prod_{\alpha=1}^{k} p_\alpha^{m_\alpha}$ and $\mathbf{t}_1 \neq \mathbf{t}_2$. In this case, we have two sub-cases.
			\begin{mysubcases}
				\item  Let $\tau = \beta \prod_{\alpha=1}^{k} p_\alpha^{m_\alpha-1}$ for some $\beta \in \{1,2, \dots, \prod_{\alpha=1}^{k} p_\alpha-1\}$. In that case $\floor{\frac{\tau}{N}}=\beta$,  $\tau \mod N = 0$ and $\Delta_N=0$. Then, form \textit{Lemma \ref{lemma_CCC}}, we have
				\begin{equation}\label{Thm_2_case_3_subcase_1_eqn_1}
					\sum_{\bm{\gamma} \in \prod_{\alpha=1}^{k} \mathbb{Z}_{p_\alpha}}^{} C \left( \psi (R_{\mathbf{t}_1}^{\bm{\gamma}} ), \psi (R_{\mathbf{t}_2}^{\bm{\gamma}} ) \right) \left(\tau \mod N \right)=0.
				\end{equation}
				\item Let $\beta \prod_{\alpha=1}^{k} p_\alpha^{m_\alpha-1} < \tau < (\beta+1) \prod_{\alpha=1}^{k} p_\alpha^{m_\alpha-1}$ for some $\beta \in \{1,2, \dots, \prod_{\alpha=1}^{k} p_\alpha-1\}$. Then $\floor{\frac{\tau}{N}}= \beta$, $0 < \tau \mod N  < \prod_{\alpha=1}^{k} p_\alpha^{m_\alpha-1}$, and  $-\prod_{\alpha=1}^{k} p_\alpha^{m_\alpha-1}< \tau \mod N-N <0 $ with $\Delta_N=1$. Then using \textit{Lemma \ref{lemma_CCC}}, we have
				\begin{equation}\label{Thm_2_case_3_subcase_2_eqn_1}
					\begin{split}
						\sum_{\bm{\gamma} \in \prod_{\alpha=1}^{k} \mathbb{Z}_{p_\alpha}}^{} C \left( \psi (R_{\mathbf{t}_1}^{\bm{\gamma}} ), \psi (R_{\mathbf{t}_2}^{\bm{\gamma}} ) \right) \left(\tau  \mod N \right)=& 0,\\
						\sum_{\bm{\gamma} \in \prod_{\alpha=1}^{k} \mathbb{Z}_{p_\alpha}}^{}  C \left( \psi (R_{\mathbf{t}_1}^{\bm{\gamma}} ), \psi (R_{\mathbf{t}_2}^{\bm{\gamma}} ) \right) \left(\tau  \mod N-N \right)= &0.
					\end{split}
				\end{equation}
			\end{mysubcases}
		\end{mycases}
		Hence, the lemma is proved.
	\end{proof}
	%
	
	We illustrate Lemma \ref{Thm_IGC_gen} by using the following example.
	\begin{example}\label{example_igc}
		Let $p_1=2$, $p_2=3$, $q=6$ and $\pi_1 (1)=1$, $\pi_2(1)=1$. Let the functions $f_1 : \mathbb{Z}_2 \rightarrow \mathbb{Q}_6$ and $f_2 : \mathbb{Z}_3 \rightarrow \mathbb{Q}_6$ be defined by
		$f_1=3v_{2,1}$ and $f_2= 4v_{3,1}$. Then the function $a_{\mathbf{s}, \mathbf{t}}^{\bm{\gamma}} : \mathbb{Z}_2 \times \mathbb{Z}_3 \times \mathbb{Z}_2 \times \mathbb{Z}_3 \rightarrow \mathbb{Q}_6$ is defined by $a_{\mathbf{s}, \mathbf{t}}^{\bm{\gamma}}= f_1 +f_2+ (3v_{2,1}\gamma_1+ 2v_{3,1} \gamma_2) + (3v_{2,1}t_1+ 2v_{3,1} t_2) + (3v'_{2} s_1+ 2v'_{3} s_2)$ where $\mathbf{s}=(s_1, s_2)$, $\mathbf{t}=(t_1, t_2)$, $\bm{\gamma}=(\gamma_1, \gamma_2)$ $\in \mathbb{Z}_2 \times \mathbb{Z}_3$. Then the code set corresponding to functions $a_{\mathbf{s}, \mathbf{t}}^{\bm{\gamma}}$ is an IGC code set of length $36$.
	\end{example}
	%
	
	\subsection{Main Construction}
	Now, we come to the main construction of 2-D ZCAC and 2-D ZCACS using EBFs obtained from the construction of the IGC code set. Before that, we state two more results which are used in the construction.
	\begin{lemma}[\cite{paterson}]\label{lemma_GCP}
		Let $q$ be an even positive integer and $f \left( x_1, x_2, \dots, x_m \right): \mathbb{Z}_2^{m} \rightarrow \mathbb{Z}_q$ be a function given by
		\begin{equation}
			f = \frac{q}{2} \sum_{\beta=1}^{m-1} x_{\pi(\beta)} x_{\pi(\beta+1)} +\sum_{\beta=1}^{m} g_\beta x_\beta +e,
		\end{equation}
		where $\pi$ is a permutation of the set $\{1,2, \dots, m\}$ and $g_\beta \in \mathbb{Z}_q$. Then the complex-valued sequences $\mathbf{a}_1$ and $\mathbf{b}_1$, corresponding to the functions 
		\begin{equation}
			\begin{split}
				a_1= &f,\\
				b_1 = &f + \frac{q}{2} x_{\pi(1)},
			\end{split}
		\end{equation}
		respectively, forms a GCP of length $2^m$ for any $e \in \mathbb{Z}_q$.
	\end{lemma}
	
	\begin{lemma}[\cite{RathinaCCC}]\label{lemma_mate}
		Let $f$, $a_1$ and $b_1$ be defined as in Lemma \ref{lemma_GCP}. Then the complex-valued sequences $\mathbf{a}_2$ and $\mathbf{b}_2$, corresponding to the functions 
		\begin{equation}
			\begin{split}
				a_2=& \bar{f} + \frac{q}{2} x_{\pi(1)},\\
				b_2=&\bar{f},
			\end{split}
		\end{equation}
		respectively, forms a GCP of length $2^m$, where
		\begin{equation}
			\bar{f}= \frac{q}{2} \sum_{\beta=1}^{m-1} \bar{x}_{\pi(\beta)} \bar{x}_{\pi(\beta+1)} +\sum_{\beta=1}^{m} g_\beta \bar{x}_\beta +e,
		\end{equation}
		and $\bar{x}_i= 1-x_i$, $\forall i$. Moreover, $(\mathbf{a}_1, \mathbf{b}_1)$ and $(\mathbf{a}_2, \mathbf{b}_2)$ are complementary mates.
	\end{lemma}
	
	For further development, we let $2 \mid q$ and $p$ be a prime such that  $p \mid q$. Also, let $a_{\mathbf{s}, \mathbf{t}}^{\bm{\gamma}}$ be the same functions used in Lemma \ref{Thm_IGC_gen}. For $\vec{d}'=(d_1, d_2, \dots, d_m, \tilde{d}_{p}) \in \mathbb{Z}_2^{m} \times \mathbb{Z}_{p}$, $\mathbf{t}_1 \neq \mathbf{t}_2$, and $\bm{\zeta}=(\mathbf{s}_1, \mathbf{s}_2, \mathbf{t}_1, \mathbf{t}_2)$, we define a set of functions
	\begin{equation}\label{func_ZCAC}
		F_{\vec{d}'}^{\bm{\gamma}, \bm{\zeta}}: \prod_{\alpha=1}^{k} \mathbb{Z}_{p_\alpha}^{m_\alpha-1} \times \prod_{\alpha=1}^{k} \mathbb{Z}_{p_\alpha} \times \mathbb{Z}_2 \rightarrow \mathbb{Q}_q,
	\end{equation}
	given by
	\begin{equation}\label{func_ZCAC_def}
		\begin{split}
			F_{ \vec{d}'}^{\bm{\gamma}, \bm{\zeta}} = &\bar{v''} \left\{a_{\mathbf{s}_1,\mathbf{t}_1}^{\bm{\gamma}} + a_1(\vec{d}) W_1 (\tilde{d}_p) + a_2( \vec{d}) W_2 (\tilde{d}_p)  \right\} \\
			&+ v'' \left\{a_{\mathbf{s}_2,\mathbf{t}_2}^{\bm{\gamma}} + b_1(\vec{d}) W _1(\tilde{d}_p) + b_2(\vec{d}) W _2(\tilde{d}_p) \right\},
		\end{split}
	\end{equation}
	where
	\begin{equation}\label{eqn_interpolation}
		\begin{split}
			W_i(\tilde{d}_p)= &\sum_{j=0}^{p-1} \frac{\prod_{i=0, i\neq j}^{p-1} (i-\tilde{d}_p)}{ \prod_{i=0, i \neq j}^{p-1} (i-j)} \delta^i_j,\\
		\end{split} 
	\end{equation}
	and
	\begin{equation}
		\begin{split}
			\delta^1_j= 
			\begin{cases}
				1 , & 2 \mid j;\\
				0, & 2 \not|~ j;
			\end{cases}& \qquad
			\delta^2_j=
			\begin{cases}
				0, & 2 \mid j;\\
				1, & 2 \not|~ j.
			\end{cases}
		\end{split}
	\end{equation} 
	Here $v''$ denotes the variable corresponding to the last $\mathbb{Z}_2$ in (\ref{func_ZCAC}), $\bar{v}'' = 1-v''$. $a_1$ and $b_1$ are the functions from Lemma \ref{lemma_GCP}, whereas  $a_2$ and $b_2$ are the functions from Lemma \ref{lemma_mate}. Also, we assumed $\vec{d}= (d_1,d_2, \dots, d_m) \in \mathbb{Z}_2^m$.
	\begin{note}
		In the function $F_{\vec{d}'}^{\bm{\gamma}, \bm{\zeta}}$, we could take the sub-function of the form (\ref{eqn_interpolation}) only after we have extended the co-domain of EBF beforehand from $\mathbb{Z}_q$ to $\mathbb{Q}_q$.
	\end{note}
	
	The only thing that now remains is to determine the values of the parameter $\bm{\zeta}$, which will produce the set size of 2-D ZCACS.
	Let us define the set $\Omega= \{ (\mathbf{s}_1, \mathbf{s}_2, \mathbf{t}_1, \mathbf{t}_2) : \mathbf{t}_1 \neq \mathbf{t}_2\}$. We take a proper subset  $\Gamma \subset \Omega$ such that for any two $\bm{\zeta}_1= (\mathbf{s}_1^1, \mathbf{s}_2^1, \mathbf{t}_1^1, \mathbf{t}_2^1), \bm{\zeta}_2= (\mathbf{s}_1^2, \mathbf{s}_2^2, \mathbf{t}_1^2, \mathbf{t}_2^2) \in \Gamma$, the terms $\mathbf{s}_i^m,\mathbf{t}_j^n$ satisfy the following condition:
	\begin{enumerate}
		\item Either, $\mathbf{t}_i^1 \neq \mathbf{t}_j^2$.
		
		\item Or, $\mathbf{s}_i^1 \neq \mathbf{s}_j^2$.
	\end{enumerate}
	Following this, we consider only those functions $\mathcal{F}_{\vec{d}'}^{\bm{\gamma}, \bm{\zeta}}$	for which $\bm{\zeta} \in \Gamma$.  Let the $\mathbb{Q}_q$-valued 2-D arrays be given by
	\begin{equation}
		\mathcal{K}^{\bm{\gamma},\bm{\zeta}} = \{ F_{ \vec{d}'}^{\bm{\gamma}, \bm{\zeta}} : \vec{d}' \in \mathbb{Z}_2^{m}  \times \mathbb{Z}_p \}
	\end{equation}
	for $\bm{\gamma} \in \prod_{\alpha=1}^{k} \mathbb{Z}_{p_\alpha}$ and we define $\mathbf{X}^{\bm{\gamma},\bm{\zeta}} = \Psi (\mathcal{K}^{\bm{\gamma}, \bm{\zeta}})$.
	Considering all these assumptions, we now propose the construction of the 2-D ZCAC and 2-D ZCACS.

	\begin{theorem}\label{thm_ZCACS}
		Each set $\mathcal{X}^{\bm{\zeta}}= \{\mathbf{X}^{\bm{\gamma},\bm{\zeta}}: \bm{\gamma} \in \prod_{\alpha=1}^{k} \mathbb{Z}_{p_\alpha}\}$ is a 2-D ZCAC of size  $L_1 \times L_2$, where $L_1= 2^m p$ and $L_2= 2 p_1^{m_1} p_2^{m_2} \dots p_k^{m_k}$, ZCZ size is $Z_1 \times Z_2$,  where $Z_1= 2^{m+1} $ and $Z_2=  p_1^{m_1-1} p_2^{m_2-1} \dots p_k^{m_k-1}$ and flock size $M= p_1 p_2 \dots p_k$. The set $\{ \mathcal{X}^{\bm{\zeta}}: \bm{\zeta} \in \Gamma\}$ is a 2-D ZCACS with set size $K= 2p_1^2 p_2^2 \dots p_k^2$.
	\end{theorem} 
	
	\begin{proof}
		To prove the theorem, we need to show that
		\begin{equation}
			\begin{split}
				\sum_{\bm{\gamma} \in \prod_{\alpha=1}^{k} \mathbb{Z}_{p_\alpha}}^{} C\left( \mathbf{X}^{\bm{\gamma},\bm{\zeta}_1}, \mathbf{X}^{\bm{\gamma},\bm{\zeta}_2} \right) (\tau_1, \tau_2)
				=
				\begin{cases}
					L_1 L_2 \prod_{\alpha=1}^{k} p_\alpha,& \bm{\zeta}_1 = \bm{\zeta}_2, (\tau_1, \tau_2 ) = (0,0);\\
					0, & \bm{\zeta}_1 = \bm{\zeta}_2, (\tau_1, \tau_2 ) \neq (0,0);\\
					& \left| \tau_1 \right| < 2^{m+1}, \\
					& \left| \tau_2 \right| < \prod_{\alpha=1}^{k} p_\alpha^{m_\alpha-1};\\
					0, & \bm{\zeta}_1 \neq \bm{\zeta}_2, \left| \tau_1 \right| < 2^{m+1},\\
					& \left| \tau_2 \right| < \prod_{\alpha=1}^{k} p_\alpha^{m_\alpha-1},\\
				\end{cases}
			\end{split}
		\end{equation}
		where $L_1=2^{m}p$ $ L_2= 2\prod_{\alpha=1}^{k} p_\alpha^{m_\alpha}$.
		It is straightforward to verify that the number of $\bm{\zeta} \in \Gamma $ is $K= 2 p_1^2 p_2^2 \dots p_k^2$. Also, for the trivial case $\bm{\zeta}_1 = \bm{\zeta}_2, (\tau_1, \tau_2 ) = (0,0)$, we have
		\begin{equation}
			\begin{split}
				\sum_{\bm{\gamma} \in \prod_{\alpha=1}^{k} \mathbb{Z}_{p_\alpha}}^{} C\left( \mathbf{X}^{\bm{\gamma},\bm{\zeta}_1}, \mathbf{X}^{\bm{\gamma},\bm{\zeta}_2} \right) (\tau_1, \tau_2)=	2^{m+1} p \prod_{\alpha=1}^{k} p_\alpha^{m_\alpha+1}.
			\end{split}
		\end{equation}
		To prove the rest of the theorem, we shall consider only $\tau_1 \geq 0$, and $0 \leq \left| \tau_2\right| < \prod_{\alpha=1}^{k} p_\alpha^{m_\alpha-1}$, because for any two 2-D arrays $\mathbf{X}_1$, $\mathbf{X}_2$, we have $C(\mathbf{X}_1, \mathbf{X}_2) (-\tau_1, \tau_2)= C(\mathbf{X}_2, \mathbf{X}_1)^* (\tau_1, -\tau_2)$. For a fixed $\bm{\gamma}$, we have
		\begin{equation}
			\begin{split}
				C\left( \mathbf{X}^{\bm{\gamma},\bm{\zeta}_1}, \mathbf{X}^{\bm{\gamma}, \bm{\zeta}_2} \right) (\tau_1, \tau_2)
				= &\sum_{i=0}^{L_1 -\tau_1-1} \sum_{j=0}^{L_2-\tau_2-1} \mathbf{X}^{\bm{\gamma}, \bm{\zeta}_1}_{i,j} {\mathbf{X}^{\bm{\gamma}, \bm{\zeta}_2}_{i', j'}}^*\\
				=& \sum_{i=0}^{L_1 -\tau_1-1} \sum_{j=0}^{L_2-\tau_2-1} \omega_q^{ 	F_{ \vec{i}}^{\bm{\gamma}, \bm{\zeta}_1} (\vec{j}) } \omega_q^{ 	-F_{ \vec{i'}}^{\bm{\gamma}, \bm{\zeta}_2} (\vec{j'}) },
			\end{split}
		\end{equation}
		where $i'=i+ \tau_1$, $j'=j+\tau_2$. Let $1+ \sum_{\alpha=1}^{k} m_\alpha= \epsilon$ and $m+1= \vartheta$. 
		We use the following notations:
		\begin{enumerate}
			\item We let $\vec{i}= (\vec{i}_{\delta}, i_{\vartheta}) $ and  $\vec{i'}= (\vec{i}'_{\delta}, i'_{\vartheta}) $, where $\vec{i}_{\delta}, \vec{i}'_{\delta}$ are the vectors with values in $\prod_{\alpha=1}^{k} \mathbb{Z}_{p_\alpha}^{m_\alpha-1} \times \prod_{\alpha=1}^{k} \mathbb{Z}_{p_\alpha}$ which are truncated from $\vec{i}$ and $\vec{i'}$, respectively.
			Whereas, $i_{\vartheta}, i'_{\vartheta}$ are the last elements of $\vec{i}, \vec{i}'$ corresponding to $\mathbb{Z}_2$.
			
			\item Similarly, we define $\vec{j}=(\vec{j}_{\xi}, j_{\epsilon})$ and $\vec{j}'=(\vec{j}'_{\xi}, j'_{\epsilon})$, where $\vec{j}_{\xi}, \vec{j}'_{\xi} \in \mathbb{Z}_2^m$ and $\vec{j}_{\epsilon}, \vec{j}'_{\epsilon} \in \mathbb{Z}_p$.  
			
			\item We define $i_{\delta}, i'_{\delta}$ to be the decimal numbers derived from the radix-representations $\vec{i}_{\delta}, \vec{i}'_{\delta}$, respectively. Similarly, statement holds for $j_{\xi}, j'_{\xi}$.
		\end{enumerate}
		Then, expanding the function $ F_{ \vec{d}'}^{\bm{\gamma}, \bm{\zeta}} (\vec{x})$, we have
		\begin{equation}\label{Thm_4_omega_eqn_1}
			\begin{split}
				\omega_q^{ F_{ \vec{i}}^{\bm{\gamma}, \bm{\zeta}_1} (\vec{j}) } \omega_q^{ -F_{ \vec{i}'}^{\bm{\gamma}, \bm{\zeta}_2} (\vec{j}') }
				=& \omega_q^{ (1-j_{\epsilon}) \left\{a_{\mathbf{s}_1^1,\mathbf{t}_1^1}^{\bm{\gamma}} (\vec{j}_{\xi}) + a_1(\vec{i}_{\delta}) W_1(i_\vartheta) + a_2(\vec{i}_{\delta}) W_2(i_\vartheta) \right\}} \\
				&  \times \omega_q^{ j_{\epsilon} \left\{a_{\mathbf{s}_2^1,\mathbf{t}_2^1}^{\bm{\gamma}} (\vec{j}_{\xi}) + b_1(\vec{i}_{\delta}) W_1(i_\vartheta) + b_2(\vec{i}_{\delta}) W_2(i_\vartheta)\right\}} \\
				& \times \omega_q^{ -(1-j'_{\epsilon}) \left\{a_{\mathbf{s}_1^2,\mathbf{t}_1^2}^{\bm{\gamma}} (\vec{j}'_{\xi}) + a_1(\vec{i}'_{\delta}) W_1(i'_\vartheta) + a_2(\vec{i}'_{\delta}) W_2(i'_\vartheta)\right\}} \\
				& \times \omega_q^{ -j'_{\epsilon} \left\{a_{\mathbf{s}_2^2,\mathbf{t}_2^2}^{\bm{\gamma}} (\vec{j}'_{\xi}) + b_1(\vec{i}'_{\delta}) W_1(i'_\vartheta) + b_2(\vec{i}'_{\delta}) W_2(i'_\vartheta)\right\}},
			\end{split}
		\end{equation}
		Now, we split the rest of the proof into three cases.
		
		\begin{mycases}
			\item Let  $0 \leq \tau_1 < 2^{m+1}$ and $0 < \tau_2 < \prod_{\alpha=1}^{k} p_\alpha^{m_\alpha-1}$. Then define
			\begin{equation}
				\begin{split}
					S_1=&  \sum_{j=0}^{\prod_{\alpha=1}^{k} p_\alpha^{m_\alpha} -\tau_2-1} \omega_q^{ 	F_{ \vec{i}}^{\bm{\gamma}, \bm{\zeta}_1} (\vec{j}) } \omega_q^{ 	-F_{ \vec{i}'}^{\bm{\gamma}, \bm{\zeta}_2} (\vec{j}') },\\
					S_2=& \sum_{j= \prod_{\alpha=1}^{k} p_\alpha^{m_\alpha} -\tau_2 }^{\prod_{\alpha=1}^{k}  p_\alpha^{m_\alpha}-1} \omega_q^{ 	F_{ \vec{i}}^{\bm{\gamma}, \bm{\zeta}_1} (\vec{j}) } \omega_q^{ 	-F_{ \vec{i}'}^{\bm{\gamma}, \bm{\zeta}_2} (\vec{j}') },\\
					S_3=& \sum_{j= \prod_{\alpha=1}^{k} p_\alpha^{m_\alpha}}^{2 \prod_{\alpha=1}^{k} p_\alpha^{m_\alpha} -\tau_2-1} \omega_q^{ 	F_{ \vec{i}}^{\bm{\gamma}, \bm{\zeta}_1} (\vec{j}) } \omega_q^{ 	-F_{ \vec{i}'}^{\bm{\gamma}, \bm{\zeta}_2} (\vec{j}') }.\\
				\end{split}
			\end{equation}
			Then, we can write the term $ C\left( \mathbf{X}^{\bm{\gamma},\bm{\zeta}_1}, \mathbf{X}^{\bm{\gamma}, \bm{\zeta}_2} \right) (\tau_1, \tau_2)$ in the following form:
			\begin{equation}\label{eqn_S_sum}
				C\left( \mathbf{X}^{\bm{\gamma},\bm{\zeta}_1}, \mathbf{X}^{\bm{\gamma}, \bm{\zeta}_2} \right) (\tau_1, \tau_2)= \sum_{i=0}^{L_1 -\tau_1-1} \bigg[ S_1 + S_2+ S_3 \bigg].
			\end{equation}
			
			\begin{enumerate}
				\item For $S_1$, we have $0 \leq j \leq \prod_{\alpha=1}^{k} p_\alpha^{m_\alpha} -\tau_2-1$. This implies that $j_{\epsilon}=j'_{\epsilon}=0$ as $\tau_2 < \prod_{\alpha=1}^{k} p_\alpha^{m_\alpha-1}$. Similarly, we can find that  $0 \leq j_{\xi} \leq \prod_{\alpha=1}^{k} p_\alpha^{m_\alpha} -\tau_2-1$,  $\tau_2 \leq j'_{\xi} \leq \prod_{\alpha=1}^{k} p_\alpha^{m_\alpha}-1$. So, using (\ref{Thm_4_omega_eqn_1}), we get
				\begin{equation}
					\begin{split}
						S_1= &\omega_q^{a_1(\vec{i}_{\delta}) W_1(i_\vartheta) + a_2(\vec{i}_{\delta}) W_2(i_\vartheta)} \omega_q^{- \left( a_1(\vec{i}'_{\delta}) W_1(i'_\vartheta) + a_2(\vec{i}'_{\delta}) W_2(i'_\vartheta) \right)}\\
						& \times \left( \sum_{j_{\xi}=0}^{\prod_{\alpha=1}^{k} p_\alpha^{m_\alpha} -\tau_2-1} \omega_q^{a_{\mathbf{s}_1^1,\mathbf{t}_1^1}^{\bm{\gamma}} (\vec{j}_{\xi})-a_{\mathbf{s}_1^2,\mathbf{t}_1^2}^{\bm{\gamma}} (\vec{j}'_{\xi}) } \right)\\
					\end{split}
				\end{equation}
				Then, whether or not $\mathbf{s}_1^1= \mathbf{s}_1^2$ or $\mathbf{t}_1^1= \mathbf{t}_1^2$, i.e., irrespective of $\bm{\zeta}_1 = \bm{\zeta}_2$ or $\bm{\zeta}_1 \neq \bm{\zeta}_2$, from \textit{Lemma \ref{Thm_IGC_gen}}, we can conclude
				\begin{equation}\label{eqn_S1}
					\sum_{\bm{\gamma} \in \prod_{\alpha=1}^{k} \mathbb{Z}_{p_\alpha}}^{} \sum_{j_{\xi}=0}^{\prod_{\alpha=1}^{k} p_\alpha^{m_\alpha} -\tau_2-1} \omega_q^{a_{\mathbf{s}_1^1,\mathbf{t}_1^1}^{\bm{\gamma}} (\vec{j}_{\xi})-a_{\mathbf{s}_1^2,\mathbf{t}_1^2}^{\bm{\gamma}} (\vec{j}'_{\xi}) }=0,
				\end{equation}
				because $0 \lneqq \tau_2 <  \prod_{\alpha=1}^{k} p_\alpha^{m_\alpha-1}$.
				
				\item For $S_2$, using a similar argument, we get
				\begin{equation}\label{eqn_S2}
					\begin{split}
						\sum_{\bm{\gamma} \in \prod_{\alpha=1}^{k} \mathbb{Z}_{p_\alpha}}^{} \sum_{j_{\xi}=\prod_{\alpha=1}^{k} p_\alpha^{m_\alpha} -\tau_2}^{\prod_{\alpha=1}^{k}  p_\alpha^{m_\alpha}-1} \omega_q^{a_{\mathbf{s}_1^1,\mathbf{t}_1^1}^{\bm{\gamma}} (\vec{j}_{\xi})-a_{\mathbf{s}_2^2,\mathbf{t}_2^2}^{\bm{\gamma}} (\vec{j}'_{\xi}) }=0.
					\end{split}
				\end{equation}
				
				\item For $S_3$, similarly we have
				\begin{equation}\label{eqn_S3}
					\begin{split}
						\sum_{\bm{\gamma} \in \prod_{\alpha=1}^{k} \mathbb{Z}_{p_\alpha}}^{} \sum_{j_{\xi}= \prod_{\alpha=1}^{k} p_\alpha^{m_\alpha}}^{2 \prod_{\alpha=1}^{k} p_\alpha^{m_\alpha} -\tau_2-1} \omega_q^{ 	F_{ \vec{i}}^{\bm{\gamma}, \bm{\zeta}_1} (\vec{j}_{\xi}) } \omega_q^{ 	-F_{ \vec{i}'}^{\bm{\gamma}, \bm{\zeta}_2} (\vec{j}'_{\xi}) }=0.
					\end{split}
				\end{equation}
			\end{enumerate}
			Now, combining (\ref{eqn_S1}), (\ref{eqn_S2}), (\ref{eqn_S3}), and (\ref{eqn_S_sum}), we have
			\begin{equation}
				\begin{split}
					\sum_{\bm{\gamma} \in \prod_{\alpha=1}^{k} \mathbb{Z}_{p_\alpha}}^{} C\left( \mathbf{X}^{\bm{\gamma},\bm{\zeta}_1}, \mathbf{X}^{\bm{\gamma},\bm{\zeta}_2} \right) (\tau_1, \tau_2)=0.
				\end{split}
			\end{equation}
			
			\item  Let $\tau_2=0$, and $0 < \tau_1 < 2^{m+1}$. Then, we have
			\begin{equation}\label{eqn_T_sum}
				\begin{split}
					C\left( \mathbf{X}^{\bm{\gamma},\bm{\zeta}_1}, \mathbf{X}^{\bm{\gamma}, \bm{\zeta}_2} \right) (\tau_1, \tau_2) =  \sum_{i=0}^{L_1 -\tau_1-1} \bigg[ T_1 + T_2 \bigg],\\
				\end{split}
			\end{equation}
			where
			\begin{equation}
				\begin{split}
					T_1= & \sum_{j=0}^{\prod_{\alpha=1}^{k} p_\alpha^{m_\alpha} -1} \omega_q^{ 	F_{ \vec{i}}^{\bm{\gamma}, \bm{\zeta}_1} (\vec{j}) } \omega_q^{ 	-F_{ \vec{i}'}^{\bm{\gamma}, \bm{\zeta}_2} (\vec{j}) },\\
					T_2= & \sum_{j= \prod_{\alpha=1}^{k} p_\alpha^{m_\alpha}}^{2 \prod_{\alpha=1}^{k} p_\alpha^{m_\alpha} -1} \omega_q^{ 	F_{ \vec{i}}^{\bm{\gamma}, \bm{\zeta}_1} (\vec{j}) } \omega_q^{ 	-F_{ \vec{i}'}^{\bm{\gamma}, \bm{\zeta}_2} (\vec{j}) }.
				\end{split}
			\end{equation}
			As $\tau_2=0$, this implies that for $T_1$ we have $j_{\epsilon}=0$, and for $T_2$ we have $j_{\epsilon}=1$.
			Hence, from (\ref{Thm_4_omega_eqn_1}), we have
			\begin{equation}\label{eqn_W1W2_1}
				\begin{split}
					T_1= &\omega_q^{a_1(\vec{i}_{\delta}) W_1(i_\vartheta) + a_2(\vec{i}_{\delta}) W_2(i_\vartheta)} \omega_q^{- \left( a_1(\vec{i}'_{\delta}) W_1(i'_\vartheta) + a_2(\vec{i}'_{\delta}) W_2(i'_\vartheta) \right)}\\
					& \times \left( \sum_{j_{\xi}=0}^{\prod_{\alpha=1}^{k} p_\alpha^{m_\alpha} -1} \omega_q^{a_{\mathbf{s}_1^1,\mathbf{t}_1^1}^{\bm{\gamma}} (\vec{j}_{\xi})-a_{\mathbf{s}_1^2,\mathbf{t}_1^2}^{\bm{\gamma}} (\vec{j}_{\xi}) } \right),\\
				\end{split}
			\end{equation}
			and
			\begin{equation}\label{eqn_W1W2_2}
				\begin{split}
					T_2=& \omega_q^{b_1(\vec{i}_{\delta}) W_1(i_\vartheta) + b_2(\vec{i}_{\delta}) W_2(i_\vartheta)}  \omega_q^{- \left( b_1(\vec{i}'_{\delta}) W_1(i'_\vartheta) + b_2(\vec{i}'_{\delta}) W_2(i'_\vartheta) \right)}\\
					& \times \left( \sum_{j_{\xi}= \prod_{\alpha=1}^{k} p_\alpha^{m_\alpha}}^{2 \prod_{\alpha=1}^{k} p_\alpha^{m_\alpha} -1} \omega_q^{a_{\mathbf{s}_2^1,\mathbf{t}_2^1}^{\bm{\gamma}} (\vec{j}_{\xi})-a_{\mathbf{s}_2^2,\mathbf{t}_2^2}^{\bm{\gamma}} (\vec{j}_{\xi}) } \right).\\
				\end{split}
			\end{equation}
			Now, we have two sub-cases.
			
			\begin{mysubcases}
				\item Let $\bm{\zeta}_1 = \bm{\zeta}_2$. In that case, we have
				\begin{equation}
					\begin{split}
						\sum_{j_{\xi}=0}^{\prod_{\alpha=1}^{k} p_\alpha^{m_\alpha} -1} \omega_q^{a_{\mathbf{s}_1^1,\mathbf{t}_1^1}^{\bm{\gamma}} (\vec{j}_{\xi})-a_{\mathbf{s}_1^2,\mathbf{t}_1^2}^{\bm{\gamma}} (\vec{j}_{\xi}) }=& \prod_{\alpha=1}^{k} p_\alpha^{m_\alpha},\\
						\sum_{j_{\xi}= \prod_{\alpha=1}^{k} p_\alpha^{m_\alpha}}^{2 \prod_{\alpha=1}^{k} p_\alpha^{m_\alpha} -1} \omega_q^{a_{\mathbf{s}_2^1,\mathbf{t}_2^1}^{\bm{\gamma}} (\vec{j}_{\xi})-a_{\mathbf{s}_2^2,\mathbf{t}_2^2}^{\bm{\gamma}} (\vec{j}_{\xi}) }=& \prod_{\alpha=1}^{k} p_\alpha^{m_\alpha}.\\
					\end{split}
				\end{equation}
				Also, using the property of $W_1$ and $W_2$, we can find that
				\begin{equation}
					\begin{split}
						W_1(i_\vartheta)=0  \Leftrightarrow  &~ W_2(i_\vartheta)=1,\\
						W_1(i_\vartheta)=1 \Leftrightarrow  &~ W_2(i_\vartheta)=0.\\
					\end{split}
				\end{equation}
				Hence, without loss of generality, from the fact that $0< \tau_1 \lneqq 2^{m+1}$ and from (\ref{eqn_W1W2_1}) and (\ref{eqn_W1W2_2}), we get
				\begin{equation}
					\begin{split}
						\sum_{i=0}^{L_1 -\tau_1-1} \bigg[ T_1 + T_2 \bigg]
						=&\prod_{\alpha=1}^{k} p_\alpha^{m_\alpha} \sum_{i_{\delta}=0}^{L_1 -\tau_1-1} \bigg[ \left( \omega_q^{ a_1 (\vec{i}_{\delta})- a_1 (\vec{i}'_{\delta})} + \omega_q^{ b_1 (\vec{i}_{\delta})- b_1 (\vec{i}'_{\delta})}  \right)\\
						& + \left( \omega_q^{ a_2 (\vec{i}_{\delta})- a_1 (\vec{i}'_{\delta})} + \omega_q^{ b_2 (\vec{i}_{\delta})- b_1 (\vec{i}'_{\delta})}  \right)\\ &+\left( \omega_q^{ a_2 (\vec{i}_{\delta})- a_2 (\vec{i}'_{\delta})} + \omega_q^{ b_2 (\vec{i}_{\delta})- b_2 (\vec{i}'_{\delta})}  \right)   \bigg]\\
						&=0,
					\end{split}
				\end{equation}
				using Lemma \ref{lemma_mate}. Therefore
				\begin{equation}
					C\left( \mathbf{X}^{\bm{\gamma},\bm{\zeta}_1}, \mathbf{X}^{\bm{\gamma}, \bm{\zeta}_2} \right) (\tau_1, \tau_2)=0.
				\end{equation}
				
				\item Let $\bm{\zeta}_1 \neq \bm{\zeta}_2$. If $(\mathbf{s}_i^1, \mathbf{t}_i^1)= (\mathbf{s}_i^2, \mathbf{t}_i^2)$, then we  proceed similar to the previous case and get
				\begin{equation}
					\begin{split}
						C\left( \mathbf{X}^{\bm{\gamma},\bm{\zeta}_1}, \mathbf{X}^{\bm{\gamma}, \bm{\zeta}_2} \right) (\tau_1, \tau_2)=0.
					\end{split}
				\end{equation}
				However, if either $\mathbf{s}_i^1 \neq \mathbf{s}_i^2$ or $\mathbf{t}_i^1 \neq \mathbf{t}_i^2$, then using Lemma \ref{Thm_IGC_gen}, we have
				\begin{equation}
					\begin{split}
						\sum_{\bm{\gamma} \prod_{\alpha=1}^{k} \mathbb{Z}_{p_\alpha}}^{ }	\sum_{j_{\xi}}^{} \omega_q^{a_{\mathbf{s}_i^1,\mathbf{t}_i^1}^{\bm{\gamma}} (\vec{j}_{\xi})-a_{\mathbf{s}_i^2,\mathbf{t}_i^2}^{\bm{\gamma}} (\vec{j}_{\xi}) }=0
					\end{split}.
				\end{equation} 
			\end{mysubcases}
			
			\item Let  $0 \leq \tau_1 < 2^{m+1}$ and $ -\prod_{\alpha=1}^{k} p_\alpha^{m_\alpha-1} < \tau_2 < 0$. This case can be proved similar to \textit{Case I}.
		\end{mycases}
		Combining all these cases, we have the theorem.
	\end{proof}
	
\begin{figure}[h]
		\centering
		\includegraphics[width=0.6\textwidth]{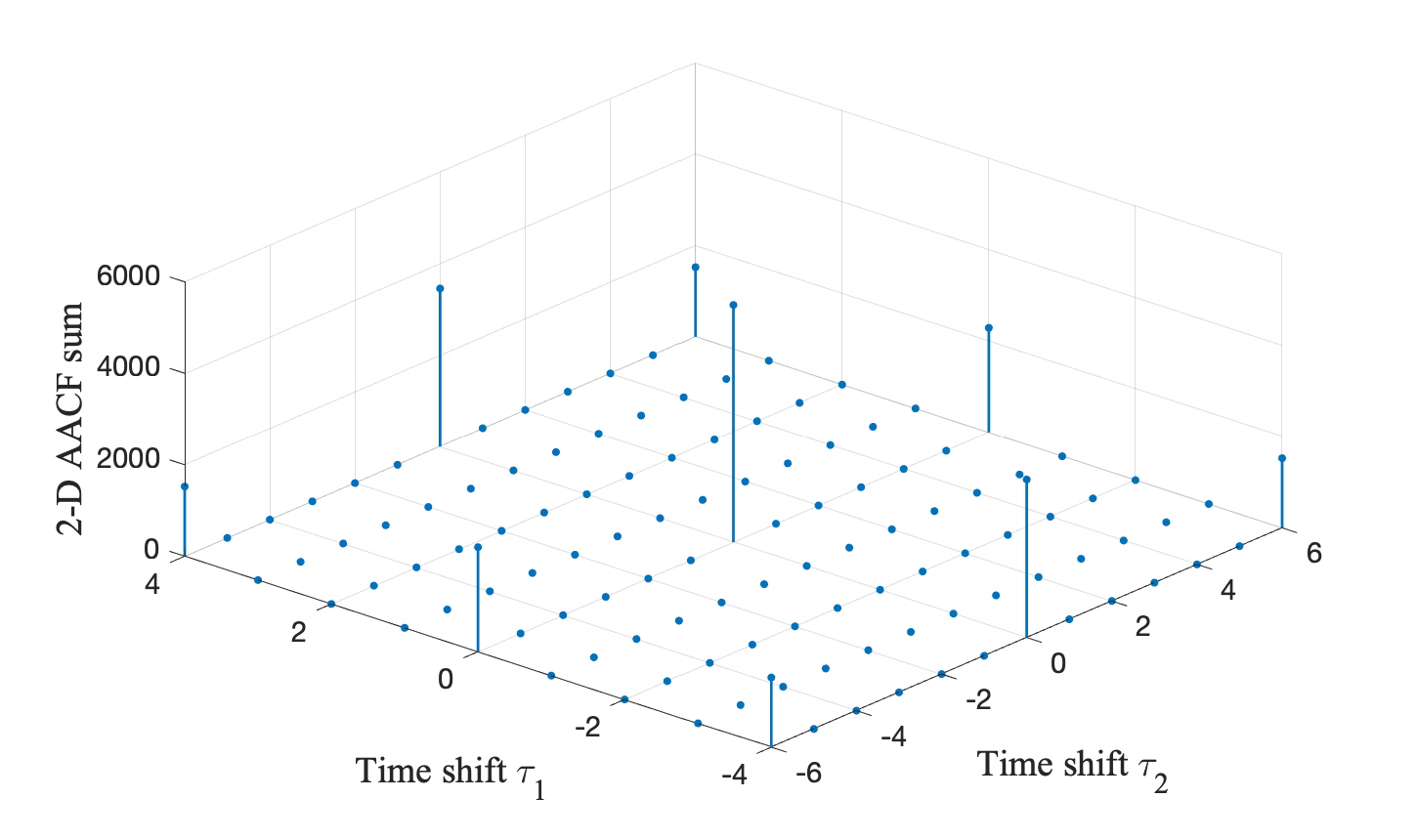}
		\caption{Plot of 2-D AACF sum corresponding to Example \ref{example_ZCACS}}\label{AACF_sum}
	\end{figure}

\begin{figure}[h]
		\centering
		\includegraphics[width=0.6\textwidth]{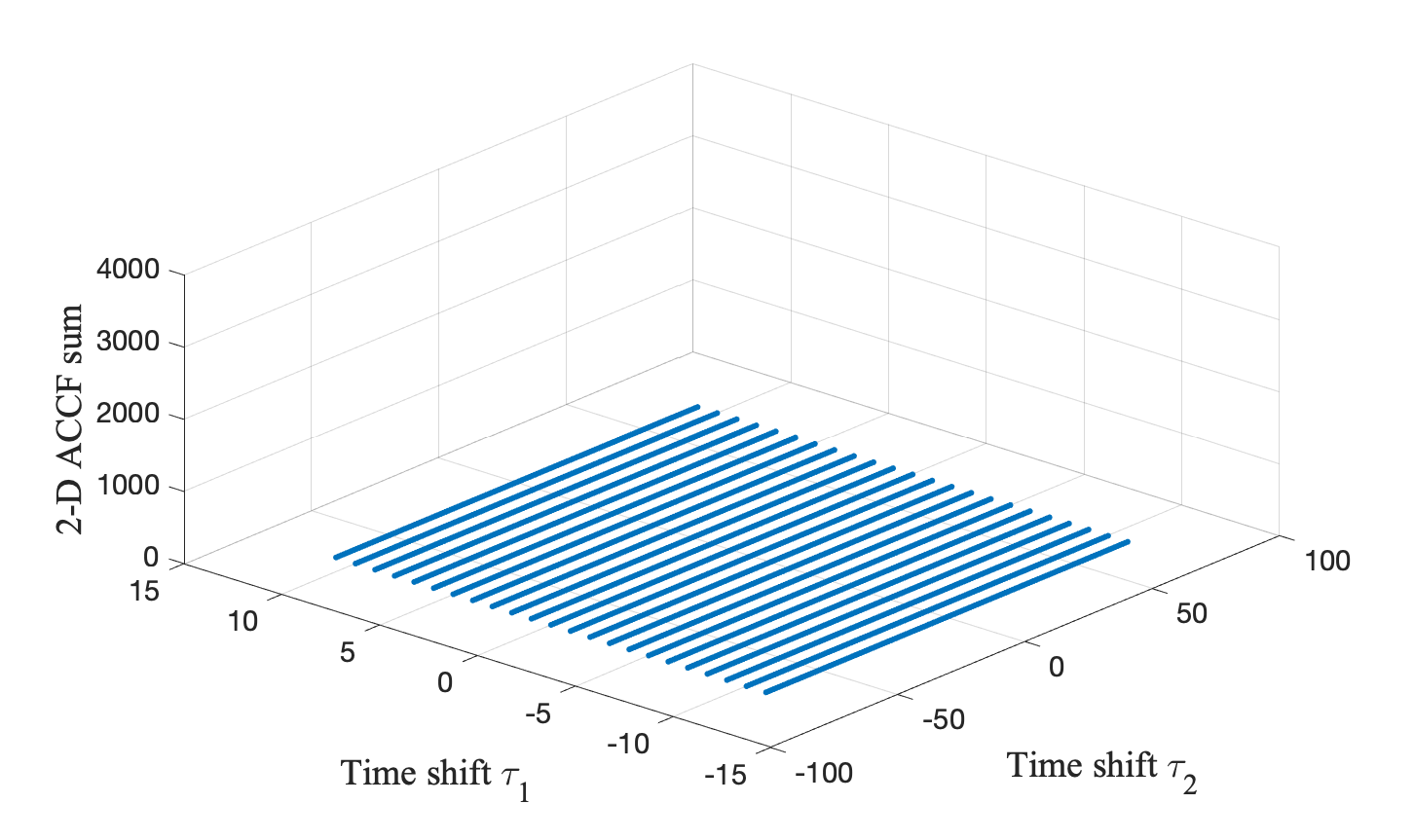}
		\caption{Plot of 2-D ACCF sum corresponding to Example \ref{example_ZCACS}}\label{ACCF_sum}
	\end{figure}

	The following example illustrates Theorem \ref{thm_ZCACS}.
	\begin{example}\label{example_ZCACS}
		Let $p_1=2$, $p_2=3$, $q=6$, $\pi_1(1)=1$,  $\pi_2(1)=1$, $a_{\mathbf{s}_i^k, \mathbf{t}_j^k}^{\bm{\gamma}}$, $f_1$,   and $f_2$ be the same functions as in Example \ref{example_igc}.
		Let the function $f: \mathbb{Z}_2^2 \rightarrow \mathbb{Z}_6$ be defined by $f = 3 x_1 x_2 +3$, where $\pi(1)=1, \pi(2)=2$. Then, let the functions $a_1,b_1,a_2, b_2: \mathbb{Z}_2^2 \rightarrow \mathbb{Z}_6$ be given by $a_1 = 3x_1 x_2+ 3$, $b_1= 3 x_1 x_2 + 3 + 3x_1$, and $a_2, b_2$ are generated from Lemma \ref{lemma_mate}. We also take $\bm{\zeta}_1= (\mathbf{s}_1^1, \mathbf{s}_2^1, \mathbf{t}_1^1, \mathbf{t}_2^1)$, where $\mathbf{s}_1^1= \mathbf{s}_2^1= (1,1)$ and $\mathbf{t}_1^1= (0,1)$, $\mathbf{t}_2^1=(1,2)$ and similarly we take $\bm{\zeta}_2= (\mathbf{s}_1^2, \mathbf{s}_2^2, \mathbf{t}_1^2, \mathbf{t}_2^2)$, where $\mathbf{s}_1^2= \mathbf{s}_2^2= (1,1)$ and $\mathbf{t}_1^2= (1,1)$, $\mathbf{t}_2^2=(1,0)$, so that $\mathbf{t}_1^1, \mathbf{t}_2^1, \mathbf{t}_1^2$ and $\mathbf{t}_2^2$ are all distinct. 
		We also take  $\vec{d}'= (d_1, d_2, \tilde{d}_p) \in \mathbb{Z}_2^2 \times \mathbb{Z}_3$, $W_1,W_2$ are generated from (\ref{eqn_interpolation}).
		Then, we get a 2-D ZCACS of size $12 \times 72$ with set size $K=6$. The 2-D AACF sum of 2-D ZCAC $\{\mathbf{X}^{\bm{\gamma}, \bm{\zeta}_1}: \bm{\gamma} \in \mathbb{Z}_2 \times \mathbb{Z}_3\}$ is given in the Fig. \ref{AACF_sum}. The size of the rectangular ZCZ is $8 \times 6$. The plot of the 2-D ACCF sum of the set $\{\mathcal{X}^{\bm{\gamma}, \bm{\zeta}_1}, \mathcal{X}^{\bm{\gamma}, \bm{\zeta}_2}\}$ is given in Fig. \ref{ACCF_sum}.
	\end{example}
	
	\begin{remark}
		It should be noted that, fixing $k=1$, and $p_1=2$, the set $\mathcal{X}= \{\mathbf{X}^\gamma : \gamma \in \mathbb{Z}_2\}$ becomes a 2-D ZCAP  $\{\mathbf{X}^0, \mathbf{X}^1\}$ of size $2^m p \times 2^{m_1 +1}$ and  rectangular ZCZ width is $2^{m+1} \times 2^{m_1 -1}$. 2-D ZCAP \cite{ChenZCAP,AbhishekZCAP} of this size is reported for the first time in this paper, as a special case.
	\end{remark}

	The next two corollaries show that we can derive 2-D GCAS and GCS from the proposed construction of 2-D ZCAC, which can be used in mMIMO system.
	\begin{corollary}\label{cor_GCAS}
		Let $\mathcal{X}^{\bm{\zeta}}$ be any 2-D ZCAC with $p=2$. Let $\bar{\mathcal{X}}^{\bm{\zeta}} $ be the set of 2-D arrays derived from $\mathcal{X}^{\bm{\zeta}}$, where each 2-D array is produced by deleting the following columns from $\mathcal{X}^{\bm{\zeta}}$: (i) From $(\prod_{\alpha=1}^{k} p_\alpha^{m_\alpha-1}+1)$-th to $(\prod_{\alpha=1}^{k} p_\alpha^{m_\alpha})$-th column; (ii) From $(\prod_{\alpha=1}^{k} p_\alpha^{m_\alpha} + \prod_{\alpha=1}^{k} p_\alpha^{m_\alpha-1}+1)$-th to $(2 \prod_{i=1}^{k} p_\alpha^{m_\alpha})$-th column. Then $\bar{\mathcal{X}}^{\bm{\zeta}} $ is a 2-D GCAS of size $(2^{m+1}) \times (2 p_1^{m_1-1} p_2^{m_2-1}  \dots p_k^{m_k-1})$ and flock size $p_1 p_2 \dots p_k $.
	\end{corollary}
	\begin{proof}
		To prove this corollary, we use Lemma \ref{Thm_IGC_gen}. Note that, after the truncation, the corresponding generating functions for the 2-D arrays of $\bar{\mathcal{X}}^{\bm{\zeta}}$ are given by
		\begin{equation}
			\begin{split}
				F_{\vec{d}'}^{\bm{\gamma}, \bm{\zeta}} = &\bar{v''} \left\{a_{\mathbf{s}_1,\mathbf{t}_1}^{\bm{\gamma}} + a_1(\vec{d}) W_1 (\tilde{d}_p) + a_2(\vec{d}) W_2 (\tilde{d}_p)  \right\} \\
				&+ v'' \left\{a_{\mathbf{s}_2,\mathbf{t}_2}^{\bm{\gamma}} + b_1(\vec{d}) W _1(\tilde{d}_p) + b_2(\vec{d}) W _2(\tilde{d}_p) \right\},
			\end{split}
		\end{equation}
		where $a_{\mathbf{s}_i,\mathbf{t}_i}^{\bm{\gamma}}= R_{\mathbf{t}_i}^{\bm{\gamma}}$, i.e., $T_{\mathbf{s}_i}=0$ for $i=1,2$. Now, proceeding similarly to the proof of Theorem \ref{thm_ZCACS}, we get the result.
	\end{proof}
	
	\begin{corollary}\label{cor_CS}
		Let $\mathbf{X}$ be any set of a 2-D ZCAC $\mathcal{X}^{\bm{\zeta}}$ with $p=2$. Then the column sequences $\{\mathbf{c}_0, \mathbf{c}_1, \dots, \mathbf{c}_{L_1-1}\}$ of $\mathbf{X}$ become a GCS of length $2^{m+1}$ and flock size $2 p_1^{m_1} p_2^{m_2} \dots p_k^{m_k}$.
	\end{corollary}
	\begin{proof}
		This is a direct consequence of the construction of 2-D ZCAC proved in Theorem \ref{thm_ZCACS}, by setting $p=2$.
	\end{proof}

	
	\section{Analysis of PMEPR Bound}\label{sec4}
	In this section, we investigate the upper bound of PMEPR for row and column sequences of the proposed 2-D ZCAC.
	
	\subsection{Row Sequence PMEPR}
	It is easy to observe that for the $m$-th row sequence $X_m^{row}$ of the 2-D array $\mathbf{X}$, we have
	\begin{equation}\label{eqn_row_PAPR}
		\begin{split}
			\left| S_{X_m^{row}} (t) \right|^2 =& L_2 + \sum_{\tau_2 \neq 0}^{} A(X_m^{row}) (\tau_2) \omega_q^{- q \tau_2 \Delta f t}.
		\end{split}
	\end{equation}
	For a fixed $\bm{\gamma}$ we take the $\mathbb{Q}_q$-valued 2-D array $K^{\bm{\gamma}}  = \{ F_{\tilde{\mathbf{d}}}^{\bm{\gamma}, \zeta} : \tilde{\mathbf{d}} \in \mathbb{Z}_2^{m} \times \mathbb{Z}_p\}$. 
	Now, from the proof of Lemma \ref{Thm_IGC_gen}, it can be deduced that
	\begin{equation}
		\sum_{\bm{\gamma} \in \prod_{\alpha=1}^{k} \mathbb{Z}_{p_\alpha}}^{}  A(K^{\bm{\gamma}}_m) (\tau_2)=0,
	\end{equation}
	for  $\beta \prod_{\alpha=1}^{k} p_{\alpha}^{m_\alpha-1} < \left| \tau_2 \right|
	< (\beta +1) \prod_{\alpha=1}^{k} p_{\alpha}^{m_\alpha-1},
	\beta= 0,1, \dots, 2\prod_{\alpha=1}^{k} p_\alpha -1$  and $\left|\tau_2 \right| = \delta \prod_{\alpha=1}^{k} p_{\alpha}^{m_\alpha-1},  \delta= \prod_{\alpha=1}^{k} p_\alpha, \prod_{\alpha=1}^{k} p_\alpha +1, \dots,2 \prod_{\alpha=1}^{k} p_\alpha-1$.
	From the above observation and the definition of AACF, we get
	\begin{equation}
		\begin{split}
			\left| \sum_{\bm{\gamma} \in \prod_{\alpha=1}^{k} \mathbb{Z}_{p_\alpha}}^{}    A(K^{\bm{\gamma}}_m) (\tau_2) \right| \leq  2 \left( \prod_{\alpha=1}^{k} p_{\alpha}- \beta \right) \prod_{\alpha=1}^{k} p_{\alpha}^{m_\alpha},
		\end{split}
	\end{equation}
	for $\left| \tau_2 \right| = \beta \prod_{\alpha=1}^{k} p_\alpha^{m_\alpha-1}$, where $\beta= 1, 2, \dots, \prod_{\alpha=1}^{k} p_\alpha-1$.
	Therefore, from (\ref{eqn_row_PAPR}), we can derive
	\begin{equation}
		\begin{split}
			\sum_{\bm{\gamma} \in \prod_{\alpha=1}^{k} \mathbb{Z}_{p_\alpha}}^{} \left|  S_{K^{\bm{\gamma}}_m} (t) \right|^2
			\leq &\sum_{ \beta=1}^{\prod_{\alpha=1}^{k} p_\alpha-1}  \sum_{ \left| \tau_2 \right| = \beta \prod_{\alpha=1}^{k} p_\alpha^{m_\alpha-1}}^{} \left| \sum_{\bm{\gamma} \in \prod_{\alpha=1}^{k} \mathbb{Z}_{p_\alpha}}^{}  A(K^{\bm{\gamma}}_m) (\tau_2) \right| \\
			&+  L_2  \prod_{\alpha=1}^{k} p_\alpha \\
			= & 2 \prod_{\alpha=1}^{k} p_\alpha^{m_\alpha+1}  \left(\prod_{\alpha}^{k} p_\alpha-1\right) + L_2 \prod_{\alpha=1}^{k} p_\alpha.
		\end{split}
	\end{equation}
	But $\left| S_{K^{\bm{\gamma}}_m} (t)  \right|^2 \geq 0, \forall \bm{\gamma} \in \prod_{\alpha=1}^{k} \mathbb{Z}_{p_\alpha}$ and $L_2 = 2\prod_{\alpha=1}^{k} p_\alpha^{m_\alpha}$. So,  we have
	\begin{equation}
		\frac{\left| S_{K^{\bm{\gamma}}_m} (t)\right|^2}{L_2} \leq \prod_{\alpha}^{k} p_\alpha \left(\prod_{\alpha}^{k} p_\alpha-1\right) + \prod_{\alpha=1}^{k} p_\alpha.
	\end{equation}
	Hence, $\text{PMEPR}(K^{\bm{\gamma}}_m) \leq \prod_{\alpha}^{k} p_\alpha  \left(\prod_{\alpha}^{k} p_\alpha-1\right) + \prod_{\alpha=1}^{k} p_\alpha $.
	Similarly, we can compute the PMEPR upper bound of the column sequence.
	

	\subsection{Column Sequence PMEPR}
	Similar to the previous case, for the $n$-th column sequence $X_n^{col}$, we have
	\begin{equation}
		\begin{split}
			\left|S_{X_n^{col}} (t) \right|^2 = L_1 + \sum_{\tau_1 \neq 0}^{} A(X_n^{col}) (\tau_1) \omega_q^{- q \tau_1 \Delta f t}.
		\end{split}
	\end{equation}
	For this, we note that either ${K^{\bm{\gamma}}}^{col}_n= \mathcal{I}_a$ or ${K^{\bm{\gamma}}}^{col}_n= \mathcal{I}_b$, where 
	\begin{equation}
		\begin{split}
			\mathcal{I}_a = &a_{\mathbf{s}_1, \mathbf{t}_1}^{\bm{\gamma}} + a_1(\vec{d}) W_1(\tilde{d}_p) + a_2(\vec{d}) W_2(\tilde{d}_p),\\ \mathcal{I}_b = & a_{\mathbf{s}_2, \mathbf{t}_2}^{\bm{\gamma}} + b_1(\vec{d}) W_1(\tilde{d}_p) + b_2(\vec{d}) W_2(\tilde{d}_p),\\
		\end{split}
	\end{equation}
	for $(\vec{d}, \tilde{d}_p) \in \mathbb{Z}_2^m \times \mathbb{Z}_p$ and $a_1, a_2, b_1, b_2, W_1$ and $W_2$ are the functions used in Theorem \ref{thm_ZCACS}. Note that for $p \neq2$, we have
	\begin{equation}
		\begin{split}
			A( \mathcal{I}_a)(\tau_1) + A( \mathcal{I}_b)(\tau_1)=
			\begin{cases}
				(p-\delta) 2^{m+1},  & \left| \tau_1 \right| = \delta 2^m, \delta = 0,2, \dots, p-1,\\
				0, & \text{otherwise}.
			\end{cases}
		\end{split}
	\end{equation}
	because $(a_2, b_2)$ is a complementary mate of $(a_1, b_1)$ by Lemma \ref{lemma_mate}, and hence
	\begin{equation}
		\begin{split}
			a_1(\vec{d}) W_1(\tilde{d}_p) + a_2(\vec{d}) W_2(\tilde{d}_p)=& \underbrace{ a_1\otimes a_2\otimes\dots\otimes a_1 }_{p\text{-times}},\\
			b_1(\vec{d}) W_1(\tilde{d}_p)+ b_2(\vec{d}) W_2(\tilde{d}_p)= &\underbrace{ b_1\otimes b_2 \otimes \dots \otimes b_1 }_{p\text{-times}}.
		\end{split}
	\end{equation}
	For $p=2$, we have 
	\begin{equation}
		\begin{split}
			a_1(\vec{d}) W_1(\tilde{d}_p) + a_2(\vec{d}) W_2(\tilde{d}_p)=& a_1 \otimes a_2\\
			b_1(\vec{d}) W_1(\tilde{d}_p)+ b_2(\vec{d}) W_2(\tilde{d}_p)=& b_1\otimes b_2.\\
		\end{split}
	\end{equation}
	In that case, $A( \mathcal{I}_a)(\tau_1) + A( \mathcal{I}_b)(\tau_1)=0, \forall \tau \neq 0$.
	So, arguing similarly to the previous case, we have
	\begin{equation}
		\text{PMEPR}({K^{\bm{\gamma}}}^{col}_n) \leq
		\begin{cases}
			2, & p=2;\\
			p+ \frac{1}{p}, & p \neq 2.
		\end{cases}
	\end{equation}
	
	\begin{remark}\label{remark_PMEPR}
		Take $p=2$, $k=1$ and $p_1=2$. Then Theorem \ref{thm_ZCACS} provides 2-D ZCAC of array size of the form $2^n \times 2^m$ and flock size $M=2$, which is comparable to the array size of 2-D GCAPs and 2-D GCASs in  \cite[Th.~12]{pai2021twodimensional}. The PMEPR of the row and column sequences of 2-D GCAP and 2-D GCAS in \cite{pai2021twodimensional} are upper bounded by $2^v$ for some positive integer $v \geq 1$. However, in our construction, for this case, the row and column sequence PMEPRs are upper bounded by $4$ and $2$, respectively. 
		This shows that the proposed 2-D ZCAC emerges as a good candidate to reduce PMEPR compared to \cite{pai2021twodimensional}.
	\end{remark}
	
\begin{figure}[h]
		\centering
		\includegraphics[width=0.6\textwidth]{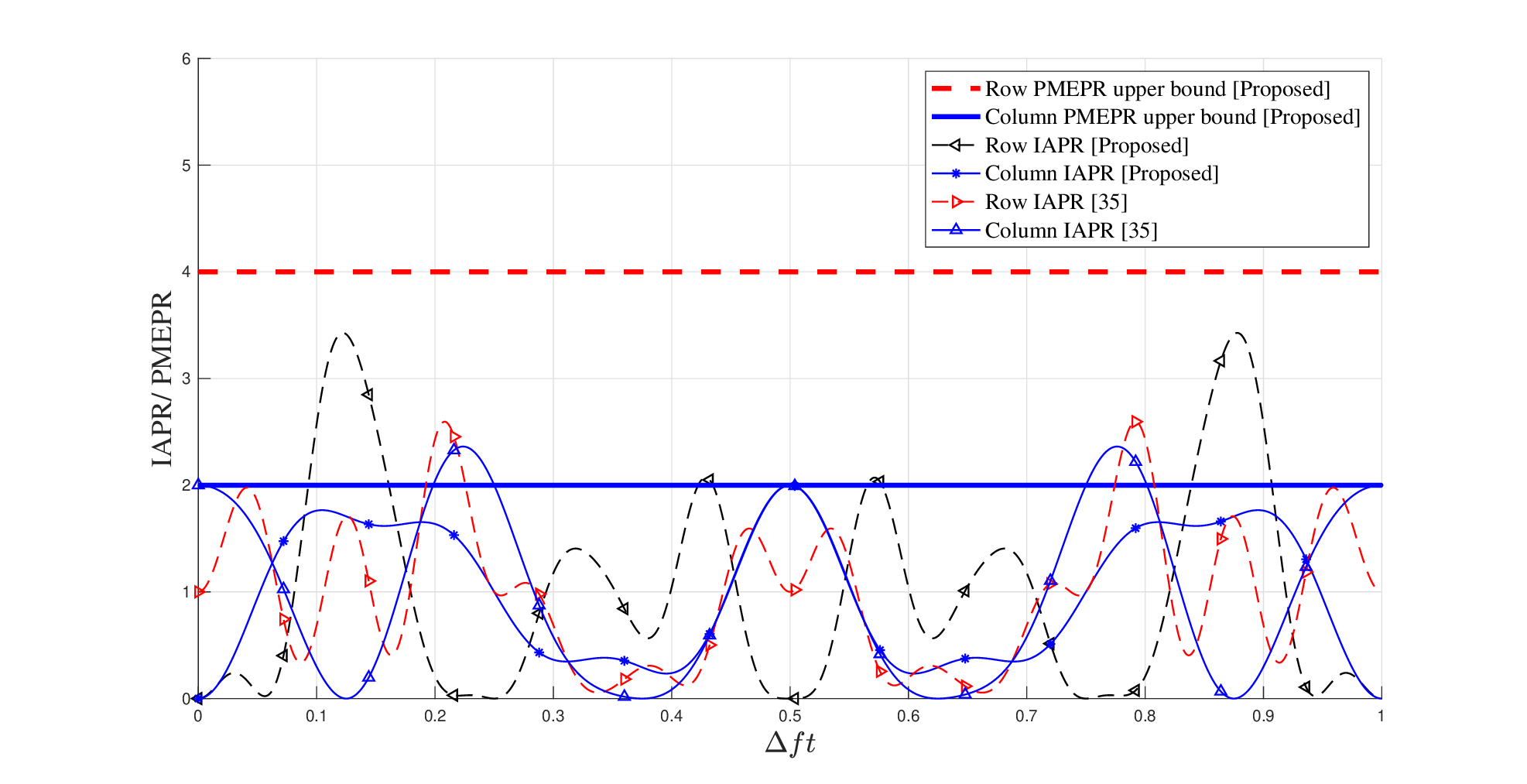}
		\caption{Row and column sequence IAPR/PMEPR}
		\label{pmepr-graph}
	\end{figure}

	Now, we verify the claim made in Remark \ref{remark_PMEPR} by an explicit example and simulation results.
	\begin{example}
		We take a 2-D ZCAC from Theorem \ref{thm_ZCACS} with the function of (\ref{func_ZCAC_def}), where $a_{\mathbf{s_i},\mathbf{t_j}}^\gamma= v_1 v_2 + \gamma v_1 + t_j  v_2+  s_i  v_3 $, $a_1= x_1 x_2 +1$, and $a_2, b_1, b_2$ are derived using Lemma \ref{lemma_GCP} and Lemma \ref{lemma_mate}.
		Then we get a 2-D ZCAC of size $8 \times 16$ and flock size $2$. The row and column sequence PMEPRs are computed to be $3.4271$ and $2$, respectively.  In Fig. \ref{pmepr-graph}, we have shown the corresponding plot of row and column sequence IAPR and PMEPR bound. We also have taken a binary $8 \times 16$-GCAP from \cite{pai2021twodimensional} by using the function $f= (z_1 z_2+  z_2 z_4+  z_4 z_5+ z_5 z_3+ z_3 z_6+ z_6 z_7)+1$ and plotted the corresponding PMEPR/IAPR curve in Fig. \ref{pmepr-graph}. It is evident from the figure that column sequence PMEPR of \cite{pai2021twodimensional} is greater than $2$, for this particular example, while the proposed one is always bounded by $2$.
	\end{example}

	\section{Bound on the Set Size}\label{sec5}
	In this section, we investigate the optimality condition of 2-D ZCACS with respect to the set size. The optimality of the code set translates to the increased number of supported users in multi-access wireless systems.	
	The existing upper bound of the set size of a $(K, Z_1 \times Z_2)$-ZCACS${}_{M}^{L_1 \times L_2}$ given in Lemma \ref{lemma_ZCACS_bound} is loose. Hence, we propose a tighter set size bound using the concept of mini-max from the paper of Welch \cite{welch} and generalizing the method presented in \cite{pai2024ZCCSbound} for ZCCS. We generalize the preliminary setting of \cite{pai2024ZCCSbound} into 2-D arrays, which are obtained from \cite{Roy2024}. For that, we state a lemma.
	\begin{lemma}[\cite{Roy2024}]\label{lemma_Roy_QCAS}
		Let $\mathcal{C}= \{\mathbf{A}^\nu: \nu= 0,1,\dots, K-1\}$ be a set of $L_1 \times L_2$ 2-D arrays with non-zero complex numbers elements of the form $\exp\left( \frac{2 \pi n\sqrt{-1}}{q}\right)$. Let 
		\begin{equation}
			c_{max}= \max\limits_{\nu_1 \neq \nu_2} \left| \sum_{i=0}^{L_1-1} \sum_{j=0}^{L_2-1} \mathbf{A}^{\nu_1}_{i,j} {\mathbf{A}^{\nu_2}}^*_{i,j} \right|.
		\end{equation}
		Then, we have
		\begin{equation}
			\begin{split}
				K (K  -1) c^2_{max} + K\left( L_1 L_2\right)^2 \geq  \sum_{s_i=0}^{L_1-1} \sum_{s_j=0}^{L_2-1} \left( \sum_{\nu=0}^{K-1}  \left| \mathbf{A}^{\nu}_{s_i,s_j} \right|^2 \right)^2. 
			\end{split}
		\end{equation}
	\end{lemma} 
	
	With this lemma, we can state the tighter set size bound for 2-D ZCACS in the following theorem.
	\begin{theorem}\label{Thm_ZCACS_bound}
		For a $(K, Z_1 \times Z_2)$-ZCACS${}_{M}^{L_1 \times L_2}$, we have the following set size bound:
		\begin{equation}
			K \leq \floor{\frac{M L_1 L_2}{Z_1 Z_2}}.
		\end{equation}
	\end{theorem}
	
	\begin{proof}
		To prove the upper bound, we adopt the concept of the 2-D cyclic shift from \cite{Roy2024}. Let $A^\nu (t)$ be the $t$-th 2-D array of the $\nu$-th 2-D ZCAC of a $(K, Z_1 \times Z_2)$-ZCACS${}_{M}^{L_1 \times L_2}$, where $t=1,2, \dots, M$. 
		We define a new set of 2-D arrays $\mathcal{D}= \{ \mathbf{D}^\nu : \nu = 0,1, \dots, K-1\}$ each having size $N_1 \times N_2$, where $N_1 = M (L_1 +Z_1-1)$ and $N_2= (L_2 +Z_2-1)$. Let $D^\nu_{i,j}$ be the $(i,j)$-th element of the 2-D array $\mathbf{D}^\nu$. Then each $\mathbf{D}^\nu$ is defined as follows:
		\begin{equation}
			D^\nu_{i,j}=
			\begin{cases}
				A^\nu_{i,j}(t), & (t-1) (L_1+Z_1-1) \leq i < (t-1) (L_1+Z_1-1)+L_1, \\
				& 0 \leq j < L_2, t \in \{1,2, \dots, M\};\\
				0, & \text{otherwise}.
			\end{cases}
		\end{equation}
		This produces $\sum_{i=0}^{N_1} \sum_{j=0}^{N_2} \left| D^\nu_{i,j} \right|^2= ML_1L_2$. 
		Now, we apply cyclic shifts on the 2-D arrays $\mathbf{D}^\nu (\nu= 1,2, \dots, K)$. We take $Z_1$ distinct cyclic shifts (including the $0$ shift) in the right direction and $Z_2$ distinct cyclic shifts (including the $0$ shift) in the downward direction and all their combinations. This procedure produces a new set $\mathcal{E}$ having $K Z_1 Z_2$ distinct 2-D arrays of size $N_1 \times N_2$. We apply \textit{Lemma \ref{lemma_Roy_QCAS}} on this set  $\mathcal{E}= \{\mathbf{E}^\nu : \nu =1,2, \dots, K Z_1 Z_2-1\}$. Then, we have
		\begin{equation}
			\begin{split}
				\overline{K}(\overline{K} - 1) c^2_{max} + \overline{K} \left( M L_1 L_2\right)^2
				& \geq  \sum_{s_i=0}^{\overline{L}_1-1} \sum_{s_j=0}^{\overline{L}_2-1} \left( \sum_{\nu=0}^{\overline{K}-1}  \left| \mathbf{E}^{\nu}_{s_i,s_j} \right|^2 \right)^2,
			\end{split}
		\end{equation}
		where $\overline{L}_1= M (L_1 + Z_1 -1)$, $\overline{L}_2= L_2 + Z_2 -1$ and $\overline{K}= K Z_1 Z_2$. But, as we are considering a 2-D ZCACS, we would have $c_{max}=0$. Hence, we get 
		\begin{equation}\label{eqn_ZCACS_condition}
			\overline{K} \left( M L_1 L_2\right)^2 \geq \sum_{s_i=0}^{\overline{L}_1-1} \sum_{s_j=0}^{\overline{L}_2-1} \left( \sum_{\nu=0}^{\overline{K}-1}  \left| \mathbf{E}^{\nu}_{s_i,s_j} \right|^2 \right)^2.
		\end{equation}
		Now, applying Cauchy-Schwarz inequality and changing the order of summation, we get
		\begin{equation}\label{eqn_ZCACS_bound_detailed}
			\begin{split}
				\sum_{s_i=0}^{\overline{L}_1-1} \sum_{s_j=0}^{\overline{L}_2-1} \left( \sum_{\nu=0}^{\overline{K}-1}  \left| \mathbf{E}^{\nu}_{s_i,s_j} \right|^2 \right)^2
				& \geq \frac{\left[ \sum_{s_i=0}^{\overline{L}_1-1} \sum_{s_j=0}^{\overline{L}_2-1} \left( \sum_{\nu=0}^{\overline{K}-1}  \left| \mathbf{E}^{\nu}_{s_i,s_j} \right|^2 \right) \right]^2}{\sum_{s_i=0}^{\overline{L}_1-1} \sum_{s_j=0}^{\overline{L}_2-1} 1}\\
				& = \frac{\left[\sum_{\nu=0}^{\overline{K}-1} M L_1 L_2 \right]^2}{M L_1 L_2} \\
				&= \overline{K}^2 M L_1 L_2.
			\end{split}
		\end{equation}
		From (\ref{eqn_ZCACS_condition}) and (\ref{eqn_ZCACS_bound_detailed}), we get
		the desired result.
	\end{proof}
	
	\begin{note}
		As the set size is always an integer, we shall call a $(K, Z_1 \times Z_2)$-ZCACS${}_{M}^{L_1 \times L_2}$ optimal, if $K = \floor{ \frac{M L_1 L_2}{Z_1 Z_2}}$.
	\end{note}	
	
	\begin{note}
		For $L_2=1$, $Z_2=1$ and setting $L_1=L$, $Z_2=Z$, we can get the set size upper bound of 1-D ZCCS proved in \cite{pai2024ZCCSbound}.
	\end{note}

	\begin{remark}
		It is straightforward to verify that the bound proposed in \textit{Theorem \ref{Thm_ZCACS_bound}} is tighter than the one derived in \cite{ZengZCACS2}. In our proposed construction, $L_1= 2^m p$, $L_2= 2 \prod_{\alpha=1}^{k} p_\alpha^{m_\alpha}$, $Z_1= 2^{m+1} $, $Z_2=  \prod_{\alpha=1}^{k} p_{\alpha}^{m_\alpha-1}$ and $M= \prod_{\alpha=1}^{k} p_\alpha$. The value of $K$ depends on the choice of $ \zeta = (\mathbf{s}_1, \mathbf{s}_2, \mathbf{t}_1, \mathbf{t}_2) \in \Gamma$ and can be calculated to be $K= 2 \prod_{\alpha=1}^{k} p_\alpha^2$. According to this newly proposed tighter upper bound, the proposed 2-D ZCACS becomes an optimal one when $p=2$.
	\end{remark}

	\section{Application in mMIMO}\label{sec6}
	Now, we discuss one more application benefit of the proposed 2-D array sets, which is mMIMO system. In Section \ref{sec3}, Corollary \ref{cor_GCAS} and Corollary \ref{cor_CS} already shows the existence of precoding matrices derived from proposed 2-D ZCACS. In this section, we provide the detailed description of the performed simulation of URA-based mMIMO transmission.
	
	\subsection{Technical Details of the Simulation}
	For the purpose of simulation, we have generated the steering matrix based on (\ref{steering}),  where we have set $\lambda = 3 \times 10^{-1} $, $d_x=\lambda/2$, $d_y=\lambda/2$ as per the convention. The received signal is obtained by using (\ref{received_signal}). We have considered only full-rate STBCs in this simulation, which have been described in \cite[Construction I]{STBC}. To compare different precoding schemes, we have fixed the URA size of the particular simulation and varied the STBC configuration because it is impossible to fix both URA size and STBC size using 2-D array codes. We compared the performance of 2-D GCAS- and GCS-based precoders from  Corollary \ref{cor_GCAS} and Corollary \ref{cor_CS}, respectively, with that of the 2-D CCC-based precoding of \cite{OmniChen}, and 2-D GCAS-based precoding of \cite{YuboGCAS,YuboGCAS2,ZilongLiuOmni2024}. The 2-D GCAPs from \cite{YuboMIMO2024} have not been considered in this simulation, as  $4 \times 6$, $8 \times 16$ and  $64 \times 54$- URA-based precoding matrices cannot be produced from the construction of \cite{YuboMIMO2024}. To decode the signals in the simulation, we have used the maximum-likelihood (ML) technique. Finally, we have obtained the BER performance and plotted it against the signal-to-noise ratio (SNR) axis, which is computed in decibels (dB). We have used randomly generated binary phase shift keying (BPSK) data bits in the simulation, and in total, $4 \times 10^{5}$ data bits are considered for each performance curve. The complex AWGN $w(t)$ used in the simulation is obtained using the formula:
	\begin{equation}
		w(t)=\sigma \left( x + \iota y \right), 
	\end{equation}	
	where $\iota=\sqrt{-1}$; $x, y \in \mathcal{N} (0,1)$ are random variables, and $\mathcal{N} (0,1)$ denotes the standard normal distribution with mean $0$, variance $1$, $\sigma^2= N_0/2$  with $N_0$ being the noise power, which is computed as $N_0= \left(  \frac{\text{Signal Power}}{\text{SNR}} \right)$. Apart from the existing 2-D GCAS/CCC-based constructions, we have also compared random matrix (RM)-based and Zadoff-Chu (ZC) sequence-based precoders \cite{Chu1972,SchroederZC}, which are widely used in practice. For RM-based precoders, the matrix elements are drawn randomly from the set $\{+1, -1\}$. For the ZC-based precoder of URA size $P \times Q$, the $K$-precoding matrices are generated by using a length-$P$ ZC sequence $\mathbf{u}$ and a length-$Q$ ZC sequence $\mathbf{v}$, with the operation
	\begin{equation}
		\mathbf{W}_n= \mathbf{u}_n \mathbf{v}_n^H, \forall n= 0,1, \dots, K-1;
	\end{equation}
	where $(\cdot)^H$ denotes the Hermitian operation, $\mathbf{u}_i, \mathbf{v}_j$ denote the $i$-th and $j$-th cyclically shifted versions of $\mathbf{u}$ and $\mathbf{v}$, respectively \cite{Omni2}. In particular, in this simulation, for both RM-based and ZC-based precoders, we set $K=4$.
	To show the performance benefit of the precoders derived from our construction, we plot simulations for three different URA sizes, including both small and large URAs, i.e., $4 \times 6$, $8 \times 16$ and $64 \times 54$-URA. Below we list all the generating functions for precoders for different URA sizes, excluding the RM-based and ZC-based precoders.
	
	For the $4 \times 6$ URA, the following precoders are used:
	\begin{enumerate}	
		\item $4\times4$-STBC is used for the precoder in \cite{ZilongLiuOmni2024}. The generating functions of $4$-ary  $\mathbf{W}_i$'s used for \cite{ZilongLiuOmni2024} is given by $f= 2 (z_1 z_2 + z_2 z_3 + z_3 z_4)+1$.
		
		\item $3\times4$-STBC is used for the derived $6$-ary 2-D GCAS-based precoding matrices. The function for 2-D GCAS is given by (\ref{func_ZCAC_def}), where $a_{\mathbf{s_i},\mathbf{t_j}}^\gamma= 2 v_1 + 2 \gamma v_1 + t_j  v_1+  2 s_i  v_2 $, $a_1= 3 x_1 + 1$ and $b_1, a_2, b_2$ follows similarly.
	\end{enumerate}
	
	For the $8 \times 16$-URA, we use the following constructions:
	\begin{enumerate}
		\item  $4\times4$-STBC is used for precoders in \cite{OmniChen, YuboGCAS}. The generating functions of binary $\mathbf{W}_i$'s used for \cite{YuboGCAS} is given by $f= (x_1 x_2 + x_2 x_3) + (y_1 y_2 + y_2 y_3 + y_3 y_4)+1$. For the case of \cite{OmniChen} also, the same $\mathbf{W}_i$'s are used.
		
		\item $8\times8$-STBC is used for the case of \cite{YuboGCAS2} with the generating function of binary $\mathbf{W}_i$'s given by $f= (x_1 x_2 + x_2 x_3) + (y_1 y_2 + y_3 y_4)+1$.
		
		\item $2\times2$-Alamouti code \cite{Alamouti} is used as STBC for the derived 2-D GCAS-based precoding matrices. The binary 2-D GCAS is constructed from \textit{Corollary \ref{cor_GCAS}}, using the function given in (\ref{func_ZCAC_def}), where $a_{\mathbf{s_i},\mathbf{t_j}}^\gamma= v_1 v_2 +v_2 v_3 + \gamma v_1 + t_j  v_3+  s_i  v_4 $, $a_1= x_1 x_2 +1$, $a_2, b_1$ and $b_2$ are derived using \textit{Lemma \ref{lemma_GCP}} and \textit{Lemma \ref{lemma_mate}}.
		
		\item The derived binary GCS-based precoding matrix is generated from \textit{Corollary \ref{cor_CS}} with the function given by (\ref{func_ZCAC_def}), where $a_{\mathbf{s_i},\mathbf{t_j}}^\gamma= v_1 + \gamma v_1 + t_j  v_1+  s_i  v_2 $, $a_1= x_1 x_2 + x_2 x_3+ 1$.  It uses the $8\times8$-STBC model given in \cite{STBC}.
	\end{enumerate}

	For the $64 \times 54$-URA system, we compare the derived 2-D GCAS with the RM-based and ZC-based precoders only, because it is not feasible to construct such precoders and STBC combinations using \cite{OmniChen,YuboGCAS,YuboGCAS2,ZilongLiuOmni2024,YuboMIMO2024}. The 2-D GCAS is derived as follows:
	\begin{enumerate}
		\item $3 \times 4$-STBC is used for the derived 2-D GCAS-based precoder. The function for generating the $6$-ary 2-D GCAS is given by $a_{\mathbf{s_i},\mathbf{t_j}}^\gamma= 2 (v_1 v_2 + v_2 v_3) + 2 \gamma v_1  + 2t_j  v_3+  2 s_i  v_4 $, $a_1= 3 (x_1 x_2 + x_2 x_3+ x_3 x_4+ x_4 x_5)+ 1$, $a_1= 3 (x_1 x_2 + x_2 x_3+ x_3 x_4+ x_4 x_5)+ 1$, and $b_1, a_2, b_2$ follows similarly.
	\end{enumerate}

	\subsection{Performance Analysis}
\begin{figure}[h]
		\centering
		\includegraphics[width=0.7\textwidth]{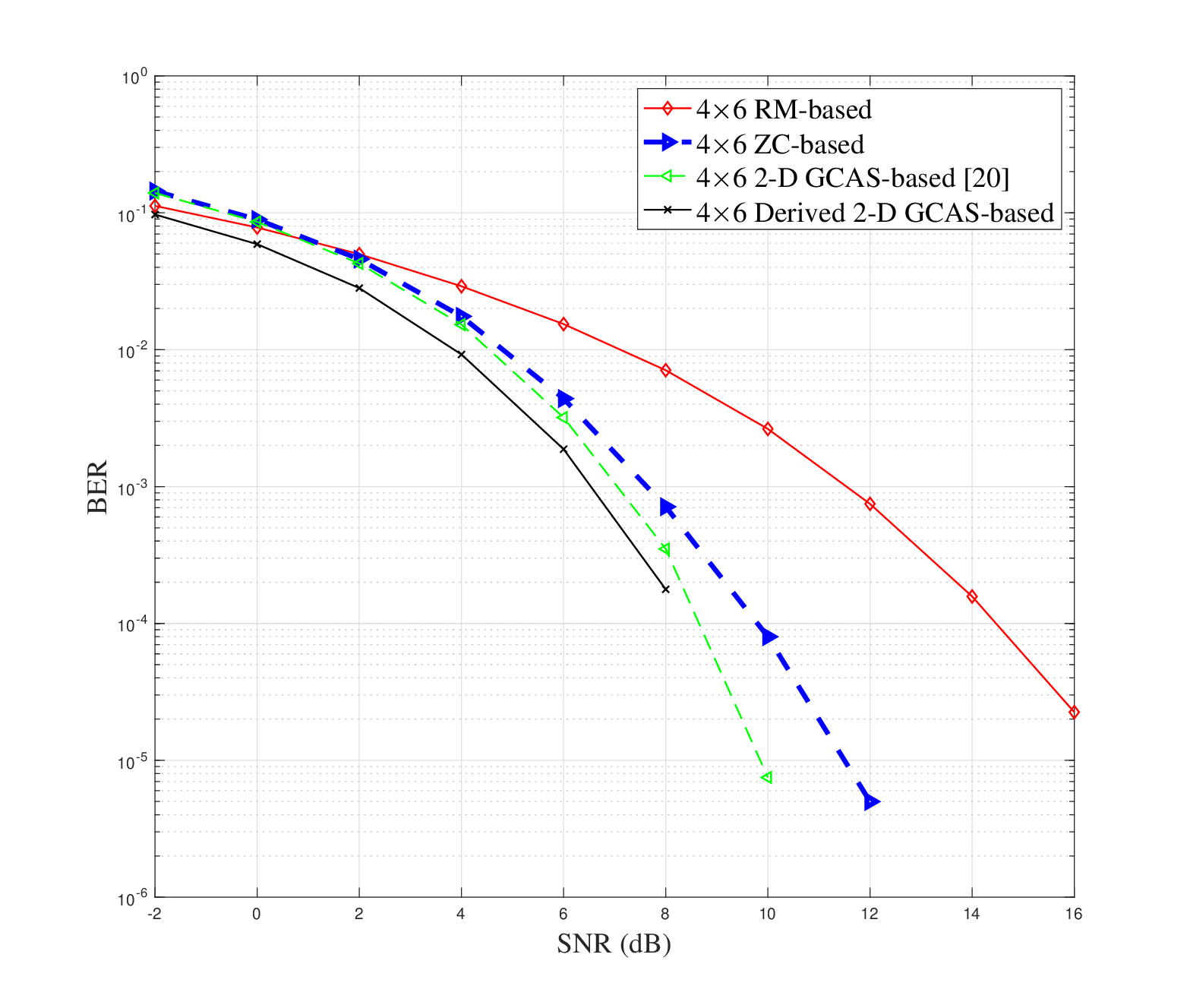}
		\caption{BER comparison for $4 \times 6$-URA}\label{fig_BER1}
	\end{figure}

\begin{figure}[h]
		\centering
		\includegraphics[width=0.7\textwidth]{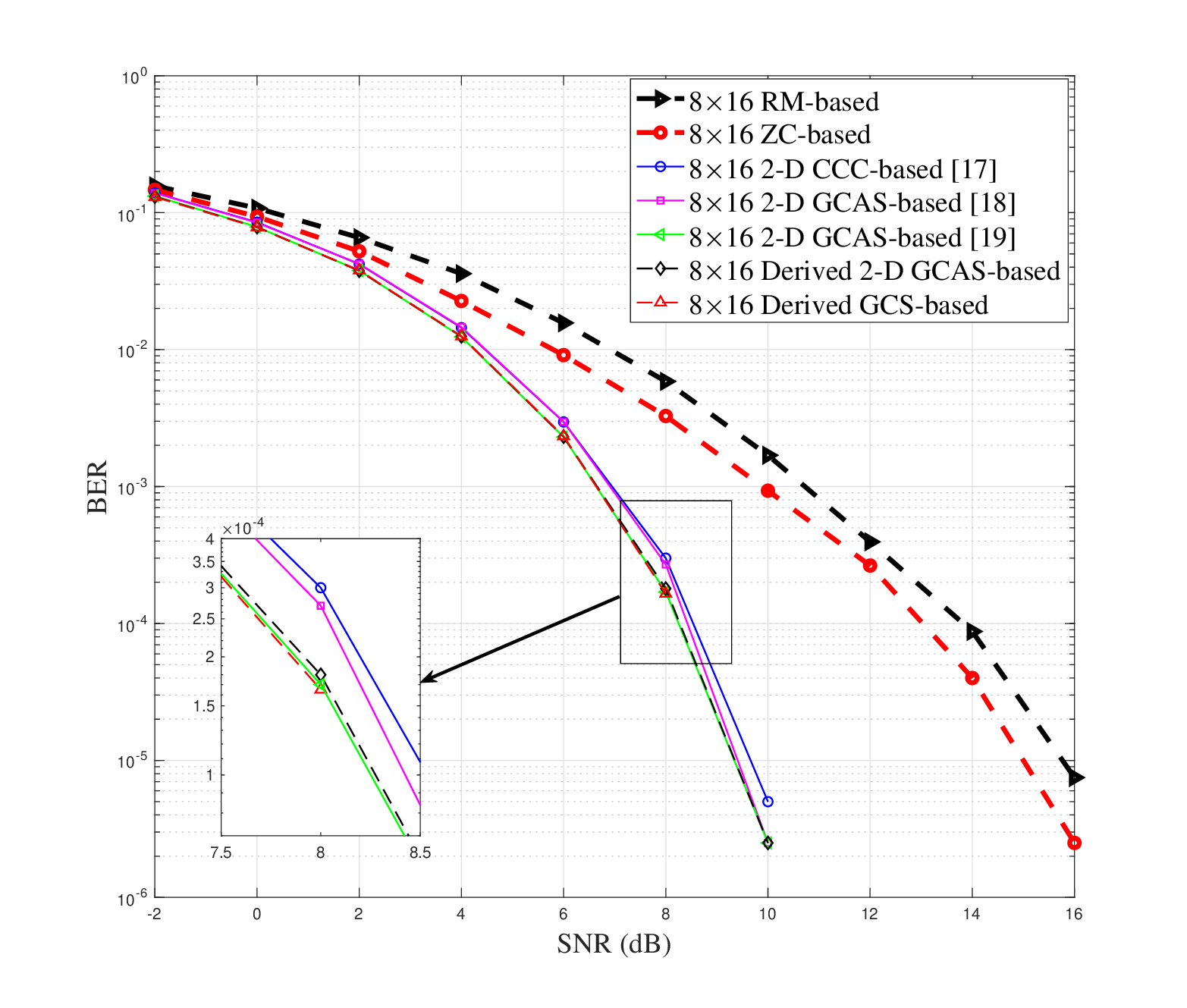}
		\caption{BER comparison for $8 \times 16$-URA}\label{fig_BER2}
	\end{figure}

\begin{figure}[h]
		\centering
		\includegraphics[width=0.7\textwidth]{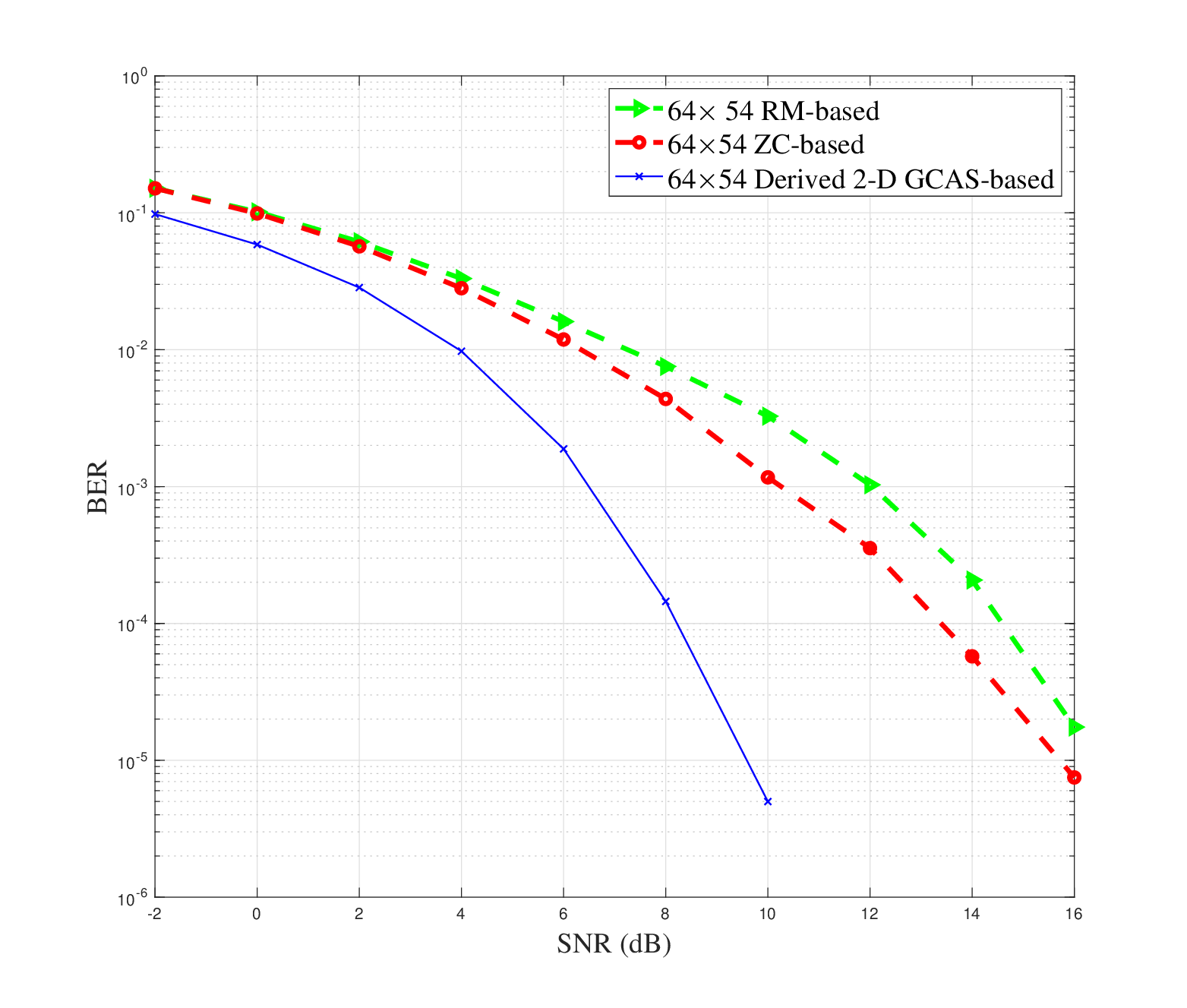}
		\caption{BER comparison for $64 \times 54$-URA}\label{fig_BER3}
	\end{figure}

	We plot all the BER performances of the above-mentioned mMIMO-URA systems with respect to varying SNR and show the application benefit of the derived 2-D arrays and codes.
	Fig. \ref{fig_BER1} clearly shows the performance benefit of the derived 2-D arrays as precoding matrix in $4 \times 6$-URA, over the existing constructions. The next figure shows the performance of precoders in larger URA size of $8 \times 16$. Fig. \ref{fig_BER2} suggests that the derived 2-D GCAS and GCS perform as good as the 2-D CCC and 2-D GCAS-based precoding in \cite{OmniChen,YuboGCAS,YuboGCAS2}, although the derived construction has benefits regarding the flexibility of the parameters. Besides that, the performances are better than widely used ZC sequence-based precoders. Note that, particularly for this example, from the point of view of complexity, the derived 2-D GCASs are more efficient than \cite{YuboGCAS2}. Because the precoder for \cite{YuboGCAS2} uses a large $8 \times 8$-STBC with the same $8 \times 16$-URA configuration increasing the transmission overhead, whereas the derived 2-D GCAS uses a small $2 \times 2$-Alamouti scheme as STBC encoder without compromising the transmit diversity. 
	Fig. \ref{fig_BER3} shows the performance benefit of the derived 2-D GCAS and GCS, when large-sized URAs are considered, $64 \times 54$-URA for this example.

	\subsection{Flexibility of System Parameters}
	The flock size of the 2-D array code determines the number of antennas to transmit STBC-encoded data. For the derived 2-D GCAS, flock size is of the form $p_1 p_2 \dots p_k$. Hence, it can support an arbitrary number of transmit antennas for STBC. Although increasing the transmit antennas for STBC encoders may result in higher complexity, here we only focus only on the flexibility theoretically. The practical applicability of antenna size for STBC encoders would depend on the system in consideration. The same theory goes for URA sizes too. From the derived 2-D arrays URA sizes can also be made flexible. Compared to \cite{OmniChen,YuboGCAS,YuboGCAS2, YuboMIMO2024,ZilongLiuOmni2024}, the derived 2-D GCAS and GCS can support flexible number of antennas in URA and STBC configurations. We show this explicitly in the next example.
	
	\begin{example}
		Consider the transmission through $3$-antenna STBC encoder with URA size $4 \times 6$ can be achieved only by the 2-D GCAS derived in \textit{Corollary \ref{cor_GCAS}}, but not by \cite{OmniChen,YuboGCAS}.
	\end{example}

	\section{Comparison with Existing Works}\label{sec7}
	In Table \ref{table_comparison_GCAS}, we compare the direct constructions of 2-D array codes used for the mMIMO system in \cite{OmniChen, YuboGCAS, YuboGCAS2, ZilongLiuOmni2024, YuboMIMO2024} with our derived 2-D GCAS and GCS, based on their parameters. Also, in Table \ref{table_comparison_ZCACS}, we compare 2-D ZCACSs in \cite{ZengZCACS2, shibsankarZCACS, Ghosh2024, YuboZCACS2025} with our proposed construction of 2-D ZCACS. It can be seen from Table \ref{table_comparison_GCAS} and Table \ref{table_comparison_ZCACS} that our proposed constructions provide a flexible set of parameters that are unreported and also can be formulated in explicit mathematical forms.
	
	\begin{table}[h]
		\centering
		\caption{Comparison table URA-STBC configuration using different 2-D arrays}
		\begin{tabular*}{\textwidth}{@{}lllll@{}}
			\toprule
			\textbf{Construction} & \textbf{Array/URA size}& \textbf{Flock size}& \textbf{\makecell{STBC required\\ (full-rate)}} & \textbf{Constraints}\\
			\midrule
			\cite{OmniChen}& $2^m \times 2^n$ & $2^k$ & $2^k \times 2^k$  & \makecell[l]{$m,n,k \geq 1$,$ m,n \geq k$}\\
			\hline
			\cite{YuboGCAS}& $p^{m_1} \times p^{m_2}$ &$p^k$ & $p^k \times N$ & \makecell[l]{$p$ is a prime, \\$m_1, m_2, k \geq 1$, \\$m_1 + m_2 \geq k$,\\ $p^k \leq \mu (N)$}\\
			\hline
			\cite{YuboGCAS2}& $b_1^{m_1} \times b_2^{m_2}$ & $N_1^{k_1} N_2^{k_2}$ & $N_1^{k_1} N_2^{k_2} \times N$  & \makecell[l]{$N_1 \geq b_1$, $N_2 \geq b_2$, \\$m_1, m_2 \geq 1$, \\$b_1, b_2 \geq  2$, \\ $k_1 \leq m_1$, $k_2 \leq m_2$, \\$N_1^{k_1} N_2^{k_2} \leq \mu ( N  )$}\\
			\hline
			\cite{ZilongLiuOmni2024}& $2^n \times N$ & $2^{k+1}$ & $2^{k+1} \times 2^{k+1} $  & \makecell[l]{$N \in \mathbb{Z}^+, n \geq 2$,\\ $k \in \mathbb{Z}^+ \cup \{0\}$}\\
			\hline
			\cite{YuboMIMO2024}& \makecell[l]{$2^{m_1} \times 10 \cdot 2^{m_2-4}$, \\$2^{m_3} \times 26 \cdot 2^{m_4-5}$}& $2$ & $2 \times 2$  & \makecell[l]{$m_1, m_3 \geq 1$, \\ $m_2 \geq 5$, $m_4 \geq 6$}\\
			\hline
			\makecell[l]{Proposed,\\
				\textit{Corollary \ref{cor_GCAS}}} & \makecell[l]{$ 2^{m+1} \times$ \\$2 p_1^{m_1-1} \cdots p_k^{m_k-1}$}& $p_1 p_2 \cdots p_k$ & \makecell[l]{$p_1 p_2 \cdots p_k \times$ \\$N$} & \makecell[l]{$m \geq 1,  m_i \geq 2, \forall i$, \\ $k \in \mathbb{N}, p_1 p_2 \cdots p_k \leq $ \\ $\mu ( N )$}\\
			\hline
			\makecell[l]{Proposed,\\
				\textit{Corollary \ref{cor_CS}}} & \makecell[l]{$ 2^{m+1} \times$ \\ $2 p_1^{m_1} \cdots p_k^{m_k}$}& $2 p_1^{m_1} \cdots p_k^{m_k}$ & \makecell[l]{$2 p_1^{m_1} \cdots p_k^{m_k} \times$ \\$ N$} & \makecell[l]{$m \geq 1,  m_i \geq 2, \forall i$,  \\$k \in \mathbb{N}, 2 p_1^{m_1} \cdots p_k^{m_k} \leq$ \\ $ \mu \left( N  \right)$}\\
			\botrule
			\multicolumn{5}{l}{\makecell[l]{  \large  \\  Note: $\mu(n)= 8c +2^d$, where $n=2^a b$,  $2 \not|~  b$, $a=4c+d$, $0 \leq d < 4$ as provided in \cite{STBC}.}}
		\end{tabular*}
		\label{table_comparison_GCAS}
	\end{table}

	\begin{sidewaystable}
		\caption{Comparison table for 2-D ZCACS}
		\begin{tabular*}{\textwidth}{@{}llllllll@{}}
			\toprule
			\textbf{Construction} &\textbf{Phase}  & \textbf{Array size} & \textbf{\makecell{Rectangular\\ ZCZ size}}& \textbf{Set size}& \textbf{Flock size} & \textbf{Constraints} & \textbf{Based on}\\
			\midrule
			\cite{ZengZCACS2} & $\{1,-1,0\}$ & $L_1 \times (L_2+ r+1)$ & $Z_1 \times Z_2$ & $K= K'r$ & 2 & \makecell[l]{$L_1, L_2 \geq 1$; $Z_1, Z_2 \geq 1$;\\ $r Z_2 \leq Z'_2$; $K \geq 1$;  \\$L_1 \times L_2$ and  $Z'_1 \times Z'_2$ \\ are the size and ZCZ \\size  of seed 2-D ZCACS \\of set size $K'$}& \makecell[l]{Matrix \\ operation} \\
			\hline	
			\cite{shibsankarZCACS} & $q$  & $L_1 \times L_2$& $Z_1 \times Z_2$  & $K$ &$M$ &\makecell[l]{$q=lcm(q_{\mathbf{V}_K}, q_{\mathbf{U}_K}, q_{\mathbf{U}_M} )$; \\$q \geq2$,  $L_1= L_2 = K$;\\ $Z_1=M, Z_2 =K$, \\$K=MP$} & \makecell[l]{Butson-type\\ Hadamard (BH) \\ matrices} \\
			\hline			
			\cite{Ghosh2024} & $q$  & $R_1 N_1 \times R_2 N_2$& $N_1 \times N_2$  & $R_1 R_2 N_1 N_2$ &$N_1 N_2$ &\makecell[l]{$R_1 \geq 1, R_2 \geq 2$, \\$N_1= p_1^{m_1}\cdots p_{k_1}^{m_{k_1}}$, \\$N_2= q_1^{n_1}\cdots q_{k_2}^{n_{k_2}}$,\\ $p_i, q_i$ are prime,\\ $m_i, n_i \geq 1$, \\$p_i \mid q$, $q_j \mid q, \forall i,j$} & EBF \\
			\hline
			\cite{YuboZCACS2025} & $q$ & $p^{m_1} \times p^{m_2}$  & $p^{m_1-s_1} \times p^{m_2-s_2}$ & $p^{k+s+1}$ &  $p^{k+1}$ & \makecell[l]{$p$ is prime, $p \mid q$, \\$k+s < m-1$, \\$k_1 +s_1 \leq m_1$, \\$k_2 +s_2 \leq m_2$} & EBF\\
			\hline
			\makecell[l]{Proposed,\\ \textit{Theorem \ref{thm_ZCACS}} } & $q$  & $L_1 \times L_2$ & $Z_1  \times Z_2$ & $2 p_1^2 p_2^2 \dots p_k^2$ & $p_1 p_2 \dots p_k$ & \makecell[l]{$L_1= 2^m p$,  \\$L_2= 2 \prod_{\alpha=1}^{k} p_\alpha^{m_\alpha}$; \\$Z_1=2^{m+1}$, \\$Z_2= \prod_{\alpha=1}^{k} p_\alpha^{m_\alpha-1} $;\\  $p_\alpha$ is prime, $\forall \alpha$; \\$p$ is prime; $m \geq 1$,  \\$m_\alpha \geq 2$, $\forall \alpha$;  \\$2 \mid q, p \mid q, p_\alpha \mid q$, $\forall \alpha$ } & EBF\\
			\botrule
		\end{tabular*}
		\label{table_comparison_ZCACS}
	\end{sidewaystable}
	
	\section{Conclusion}\label{sec8}
	In this paper, we first constructed IGC code set as a foundation, and using it 2-D ZCAC and 2-D ZCACS with flexible array sizes have been proposed. The bounds of row and column sequence PMEPRs have been investigated and shown to have advantages over existing constructions. 
	We have also proposed a new upper bound for the set size of 2-D ZCACS. The proposed construction is optimal with respect to the theoretical upper bound for special cases. In addition to that, 2-D GCAS and GCS are derived from the proposed 2-D ZCAC and applied as precoding matrices in the mMIMO systems with URA-STBC for omnidirectional transmission. Simulation results show that the derived 2-D arrays have better BER performance than the ones in practice. Also, they offer a wide range of flexibility in terms of URA size and the number of antennas for STBC encoding. It is an interesting research problem for future to investigate the construction of 2-D ZCACS with tighter PMEPR upper bound, and more parameter flexibility with optimal set size.
	
	\backmatter

	

	\bibliography{Abhishek_ZCACS_ref}

\end{document}